\documentclass[aps,prb,reprint,groupedaddress, superscriptaddress]{revtex4-2}

\usepackage{graphicx}
\usepackage{amssymb}
\usepackage{amsmath}
\usepackage{bm}
\usepackage{hyperref}
\usepackage{lipsum}
\usepackage{bbold}
\usepackage{comment}
\usepackage{xcolor}
\usepackage{longtable}

\begin{document}

\title{Commensuration torques in double-moir\'e twisted trilayer hexagonal boron nitride and graphene heterostructures} 

\author{Youngju Park}
\affiliation{Department of Physics, University of Seoul, Seoul 02504, Korea}
\affiliation{SKKU Advanced Institute of Nanotechnology and Department of Nano Engineering, Sungkyunkwan University, Suwon 16419, Korea}

\author{Nicolas Leconte}
\altaffiliation[Present address: ]{Catalan Institute of Nanoscience and Nanotechnology (ICN2), CSIC and BIST, Campus UAB, Bellaterra, 08193 Barcelona, Spain.}
\affiliation{Department of Physics, University of Seoul, Seoul 02504, Korea}

\author{Prathap Kumar Jharapla}
\affiliation{Department of Physics, University of Seoul, Seoul 02504, Korea}

\author{Md Shaifullah}
\affiliation{Department of Physics, University of Seoul, Seoul 02504, Korea}

\author{E. H. Hwang}
\email[Contact author: ]{euyheon@skku.edu}
\affiliation{SKKU Advanced Institute of Nanotechnology and Department of Nano Engineering, Sungkyunkwan University, Suwon 16419, Korea}

\author{Jeil Jung}
\email[Contact author: ]{jeiljung@uos.ac.kr}
\affiliation{Department of Physics, University of Seoul, Seoul 02504, Korea}

\begin{abstract}
We study commensuration-driven torques and angle locking in double-moir\'e trilayer hexagonal boron nitride (hBN) and graphene heterostructures using large-scale atomistic relaxations. 
In twisted trilayer hBN (t3BN) homostructures, double-moir\'e commensuration ($\theta_{12} = -\theta_{23}$) give rise to local energy minima accompanied by torque sign reversals, signaling a restoring tendency toward the commensurate configuration. 
The corresponding binding energies are $\sim$0.2-0.3 meV/atom, originating from enhanced overlap of low-energy stacking domains, although the system is globally stable at zero twist.
In contrast, in graphene/hBN heterolayers systems the global energy minimum can coincide with the double-moir\'e commensuration angle, particularly near $\sim$0.6$^{\circ}$, reflecting competition between lattice mismatch and interfacial relaxation. Incommensurate atomic structures have reduced stabilization due to suppressed overlap of low-energy stacking and have enhanced superlubricity due to spatial averaging of interfacial energies.
These results establish double-moir\'e commensuration as a general, system-dependent mechanism for twist-angle stabilization, whose angular stability is characterized by the torque magnitude and binding energy.
Coulomb electrostatic interactions further enhance the stabilization energy without changing the underlying physics.
\end{abstract}

\maketitle

\section{Introduction}
Research has been shifting toward supermoir\'e systems that can be realized by stacking two or more moir\'e interfaces in structures with at least three layers~\cite{Finney2019, Wang2019nl, Anelkovi2020, PhysRevB.81.125427, Wang2019bis, PhysRevLett.127.166802, Leconte2020,Lee2020}, going beyond single moir\'e systems obtained by combining two two-dimensional (2D) layered materials with a twist angle and/or lattice mismatch between the layers~\cite{Dean:bv, Geim2013}. 
Recent work on double moir\'e systems, including twisted trilayer graphene (t3G) and hBN-aligned twisted bilayer graphene (t2G/hBN), predicted two notable effects~\cite{leconte2023commensuration}:
(i) moir\'e superlubricity, already present in single-moir\'e structures~\cite{
doi:10.1021/acsanm.1c01540,
Bai2022,
PhysRevLett.92.126101,
Wang2019,
doi:10.1021/nl204547v,
PhysRevLett.100.046102,
1908.04666,
PhysRevLett.111.235502,
PhysRevB.70.165418,
TANG2023108288,
WangKQ2019}, manifests similarly in the \textit{incommensurate} regime of supermoir\'e systems, and
(ii) this superlubric behavior is interrupted 
at specific twist angles by potential energetic locking 
when the system transitions from \textit{incommensurate} supermoir\'e phases to the \textit{commensurate} supermoir\'e configurations.
There, the \textit{commensurate} configurations are defined as supermoir\'e systems in which the two moir\'e lattices have the same lattice constants and orientations, up to integer multiples of 60$^{\circ}$.
In the \textit{incommensurate} phases, the top layer is twisted away from this \textit{commensuration}, resulting in two distinct moir\'e length scales.
Because the twist angle serves as a continuous control parameter, the associated energetics can be naturally analyzed via the torque, defined as the angular derivative of the total energy with respect to the twist angle. A sign reversal of the torque across the \textit{commensurate} angle then signals a restoring tendency toward the locked configuration, with its magnitude reflecting the strength of the angular stability.

The mechanism behind this angle-locking effect is expected to be rather universal.
Thus, in this work we perform a systematic survey of the most common trilayer double moir\'e combinations of graphene (G) and hBN not considered in earlier. 
We begin by calculating the sliding- and twist-angle-dependent energetics for 3-layer combinations of hBN (t3BN), where the inequivalence between the B and N atoms leads to several dissimilar stacking sequences compared to trilayer graphene systems. 
We note that twisted 2-layer hBN (t2BN) systems have been extensively studied thanks to their nearly flat bands and local ferroelectric charge transfer between layers~\cite{xian2019multiflat,yao2021enhanced,vizner2021interfacial,yasuda2021stacking,woods2021charge,moore2021nanoscale,ni2019soliton,walet2021flat,zhao2020formation}. 
We then explore heterolayer systems with lattice constant mismatch by combining both G and hBN such as hBN-encapsulated graphene and graphene on t2BN.
The former system has been shown to host a variety of moir\'e-of-moir\'e signatures such as higher-order Brown-Zak oscillations, minigap formation, and constructive/destructive interference in superlattice signatures~\cite{Wang2019bis,Wang2019nl, Leconte2020, Anelkovi2020,Lee2020}, while the latter system may be promising for studying proximity-induced electrostatic charging due to t2BN ~\cite{vizner2021interfacial}. 
Although the general conclusions from our previous t3G work remain broadly applicable for the new systems considered, we identify system-specific behavior in these G/hBN-based systems that may be directly relevant for experimental efforts aiming to leverage angle locking in double-moir\'e structures.
One notable observation that contrasts with our results in t3G is that   
the double-moir\'e \textit{commensurate} structures, which normally give rise to  local energy minima, can become the globally most stable configurations in certain systems.
For t3G the global energy minimum was found when the top two layers are perfectly aligned, with the zero twist angle between them, leading to a single moir\'e structure.

The manuscript is organized as follows. The calculation methodology is introduced in Sect.~\ref{methodSect}. In Sect.~\ref{sec:results}, we present our results in Sect.~\ref{sec:onlyBN} we summarize our findings for all t3BN allotropes, and in Sect.~\ref{sec:BNandG} we turn to the heterolayer systems where G and hBN are concurrently present. Conclusions are finally drawn in Sect.~\ref{sec:conclusions}. 

\begin{figure*}[tbhp]
\begin{center}
\includegraphics[width=0.8\textwidth]{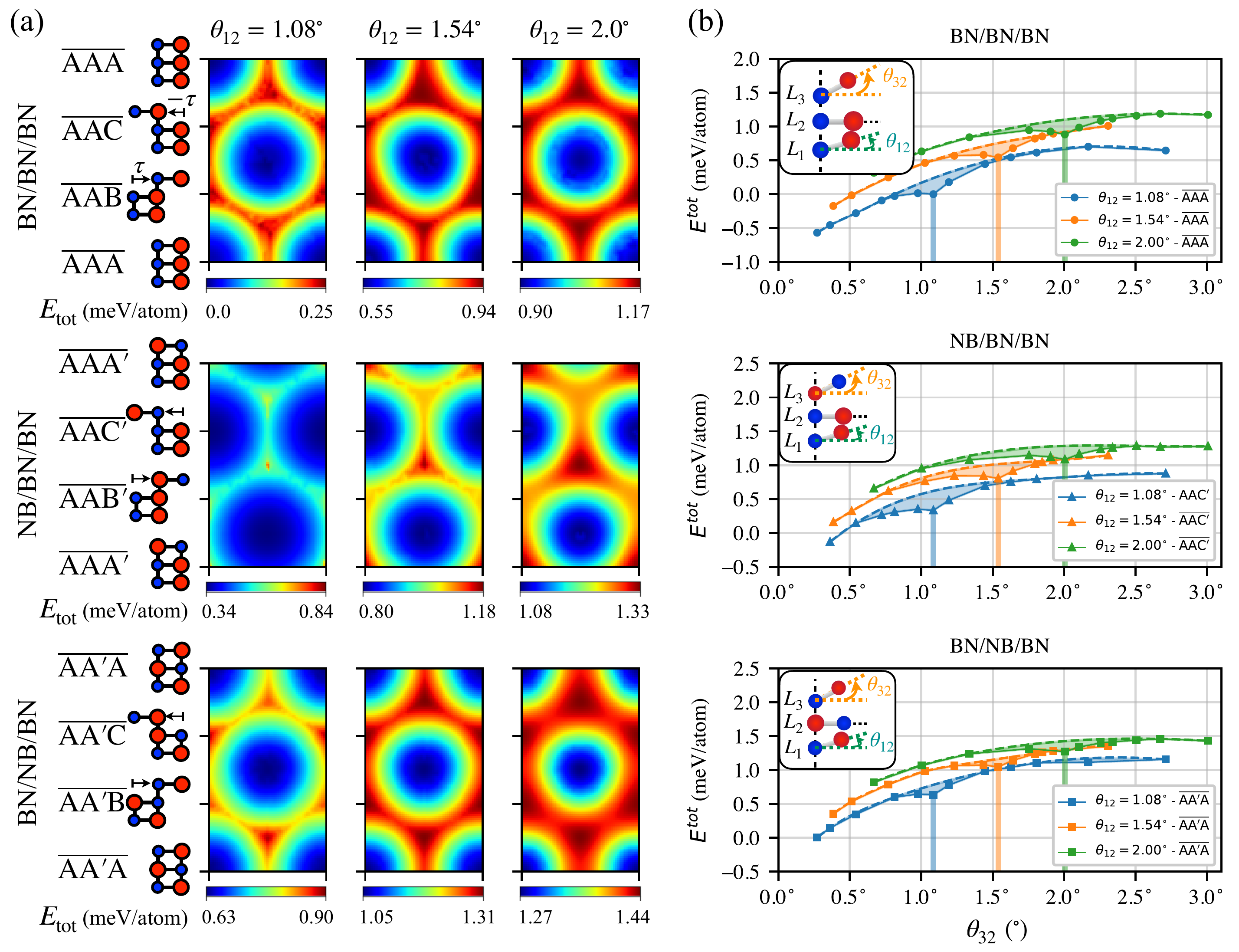}
\caption{(color online)
Sliding- and twist-angle-dependent total energies per atom ($E_{tot}$) for t3BN systems. 
(a) Sliding-dependent energy maps for double-moir\'e \textit{commensurate} structures at $\theta_{12}=\theta_{32} = 1.08^\circ$, $1.54^\circ$ and $2.00^\circ$ for different relative orientations between the hBN layers.
When sliding the top layer, the BN/BN/BN, NB/BN/BN, and BN/NB/BN structures are most stable for the $\overline{\text{AAA}}$, $\overline{\text{AAC}^\prime}$, and $\overline{\text{AA}^\prime\text{A}}$ configurations, respectively. 
The overline notation on a stacking label denote a configuration in which that local stacking is placed at the rotation center before applying the twist. 
Among these three, BN/BN/BN with the ${\text{AAA}}$ rotation center has the lowest total energy.
(b) Angle-dependent transition from \textit{incommensurate} to \textit{commensurate} configurations.
Total energies per atom are shown for fixed values of $\theta_{12} = 1.08^\circ$, $1.54^\circ$ and $2.0^\circ$ between the bottom two layers, as a function of $\theta_{32}$ for the top two layers, as schematically defined in the inset. 
(inset) Schematic illustrating the different t3BN systems and defining the twist angles of layer $i$ with respect to layer $j$ as $\theta_{ij} = \theta_i - \theta_j$. 
A local minimum is observed for the \textit{commensurate} configurations where $\theta_{12}=\theta_{32}$, which are indicated by vertical lines. 
All total energies per atom are globally shifted by a constant value of 6.7151 meV/atom, such that the $E_{tot}$ of BN/BN/BN with the $\overline{\text{AAA}}$ configuration at $\theta_{12}=\theta_{32} = 1.08^\circ$ is set to zero. 
}
\label{fig:energy}
\end{center}
\end{figure*}

\section{Methodology}
\label{methodSect}

The methodology to obtain the atomic structures follows closely Ref.~\cite{leconte2023commensuration}.
For the atomic relaxation, we minimize the total energy 
\begin{eqnarray}
E_{\rm tot} = \frac{\left[\sum_{\text{layer}\ \ell} \sum_{i\in \text{layer}\ \ell} \frac{1}{2}\left( E_{\rm el}^{i} + E_{\rm pot}^{i} \right)\right]+E_{\rm corr}}{N_{\rm tot}+N_{\rm corr}}
\label{eq:tot}
\end{eqnarray}
using LAMMPS~\cite{Plimpton1995} with the FIRE minimization scheme~\cite{PhysRevLett.97.170201} with a time step of $0.001$~ps and a stopping tolerance on the forces of $0.001~\text{eV/\AA}$. 
The elastic (el) and potential (pot) energy components for an atom $i$ are written as
\begin{equation} \begin{aligned}
E_{\rm el}^{i\in\text{layer}\ell}  &= \sum_{j\in\text{layer}\ell} \phi_{\rm intra}^{ij} \\
E_{\rm pot}^{i\in\text{layer}\ell} &= \sum_{j\notin\text{layer}\ell} \phi_{\rm inter}^{ij}
\end{aligned}\label{eq:el_pot}\end{equation}
using the extended Tersoff potential (BN-ExTeP)~\cite{PhysRevB.96.184108} and the REBO2 potential (CH.rebo)~\cite{Brenner_2002} for the intralayer interaction $\phi_{\rm intra}^{ij}$ of hBN and graphene, respectively, with the equillibrium lattice constant of $a_{\rm G}=2.46019~\text{\AA}$ and $a_{\rm BN}=2.504~\text{\AA}$, and the EXX-RPA-informed~\cite{PhysRevB.96.195431} DRIP~\cite{PhysRevB.98.235404} parametrizations for the interlayer force-fields $\phi_{\rm inter}^{ij}$.
The \textit{commensurate} supercells, as well as large rational approximants used for \textit{incommensurate} twist angles, are built using six integers enforcing commensurability conditions, as recalled in Appendix~\ref{HermannSection}.
The correction terms $N_{\rm corr}$ and $E_{\rm corr}$ in Eq.~(\ref{eq:tot}) compensate for the artificial variation of the atom-number-averaged total energy arising from changes in the graphene-to-hBN atom ratio.
This artifact originates from the different reference energies per atom of graphene and hBN.
To remove this artifact, we define the correction terms such that the layer-resolved atom-number ratio remains identical to that of the corresponding double-moir\'e \textit{commensurate} structure,
\begin{equation}
N_{\rm corr}
\equiv
N_{\rm L_3}^{\rm ref}
-N_{\rm L_3},
\qquad
E_{\rm corr}
\equiv
E_{\rm L_3}^{\rm ref} N_{\rm corr},
\end{equation}
where
\begin{equation}
N_{\rm L_3}^{\rm ref}
=
\left(
\frac{\lambda}
{\lambda_{\rm comm}}
\right)^2
N_{\rm L_3}^{\rm comm}.
\end{equation}
Here, $E_{\rm L_3}^{\rm ref}$ is taken as $-7.394$~eV/atom for a graphene top layer and $-6.690$~eV/atom for a hBN top layer, evaluated at their equilibrium lattice constants $a_{\rm G}$ and $a_{\rm BN}$, respectively~\cite{GBN}.

To clearly isolate the double-moir\'e effects, 
we fix the bottom-layer rotation angle ($\theta_{12}$) by maintaining the ratio between the first four integers, and vary the top-layer rotation angle $\theta_{32}$, which serves as a continuous control parameter. 
We then consider the angular derivative of the total energy with respect to $\theta_{32}$, 
$k(\theta_{32})\equiv {d E_{tot}}/{d \theta_{32}}$,
which corresponds to the torque acting on the top layer and reflects a restoring tendency toward energetically favorable configurations.
A sign reversal of this torque across the double-moir\'e \textit{commensuration} angle provides direct evidence of angle locking. 
To quantify this behavior, we define the one-sided torque constants on the left ($-$) and right ($+$) sides of the \textit{commensurate} angle $\theta_{32}^{\text{comm.}}$ as
\begin{equation}
k_{\pm}(\theta_{32}) \equiv 
\left.\frac{d E_{\text{tot}}}{d \theta_{32}}\right|_{\theta_{32}=\theta_{32}^{\text{comm.}}\pm 0}
\end{equation}
The magnitude of $k_{\pm}(\theta_{32})$ provides a local measure of the angular restoring strength, while the binding energy $E_{b}$ quantifies the associated energy gain relative to a smooth reference curve.
Here, the reference curve is obtained by cubic-spline interpolation from 3--5 data points selected on each side of the \textit{commensuration}-induced dip, away from its immediate vicinity.

\section{Results}
\label{sec:results}

We first consider homolayer systems with equal lattice constants (Sect.~\ref{sec:onlyBN}), as in t3G, and then address the heterolayer systems (Sect.~\ref{sec:BNandG}). Schematic insets in the main figures illustrate the corresponding systems.

\subsection{Equal lattice constant homolayer systems}
\label{sec:onlyBN}

\begin{figure}[tbhp]
\begin{center}
\includegraphics[width=1.0\columnwidth]{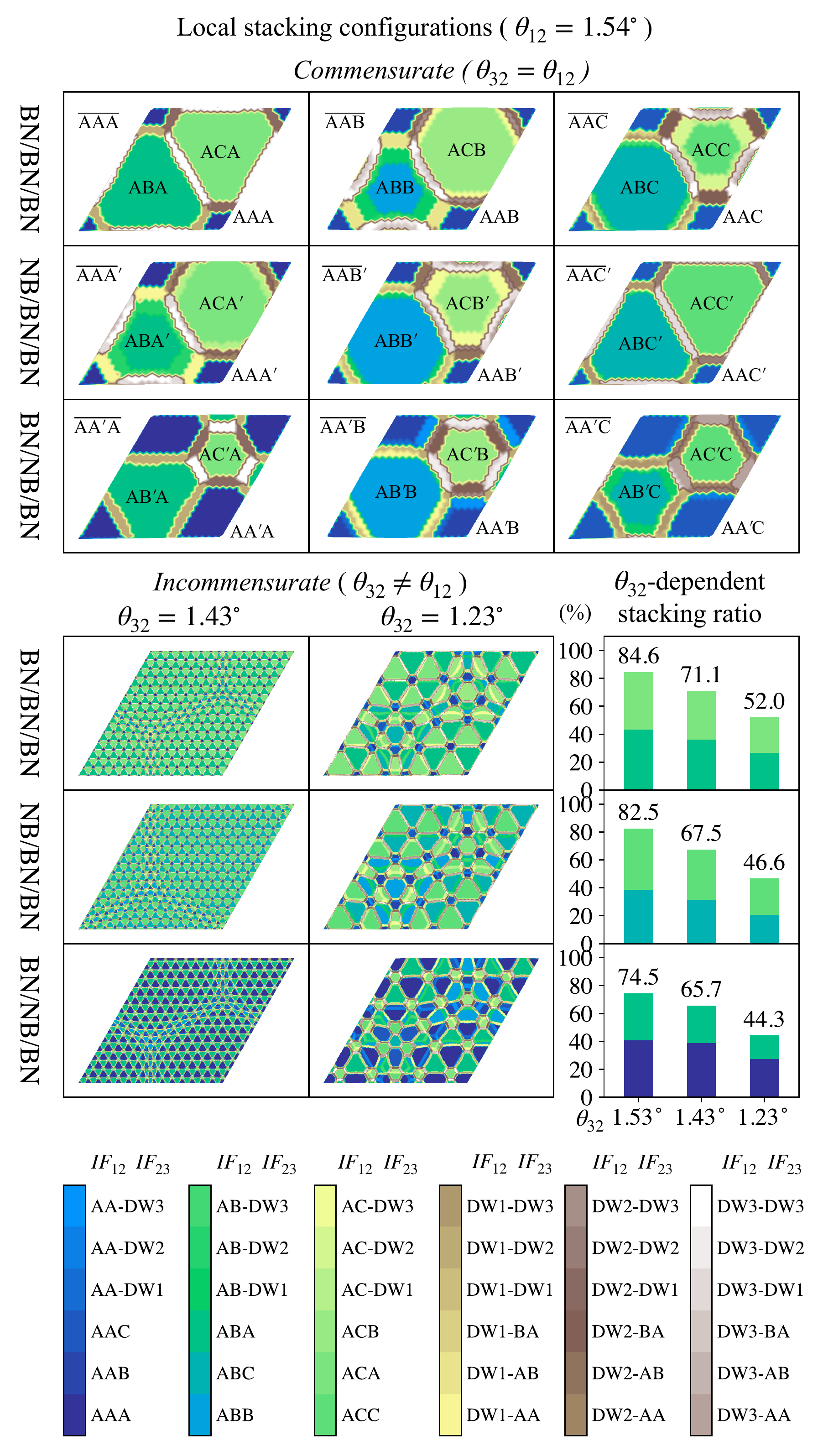}
\caption{(color online)
Sliding-dependent local stacking configurations for both \textit{commensurate} and \textit{incommensurate} t3BN systems are shown for $\theta_{12}=1.54^{\circ}$. 
The local stacking configurations after atomic relaxation are presented by color maps, which are divided into 36 sectors corresponding to local-stacking domains defined by the combined relative stackings at the two moir\'e interfaces, $IF_{12}$ and $IF_{23}$, between the middle/bottom and top/middle layers, respectively.  
For each interface, we adopt the six-sector classification of Ref.~[\onlinecite{PhysRevB.106.115410}], consisting of three high-symmetry stacking configurations (AA, AB, BA (or AC)) and three intermediate domain walls (DW1, DW2, DW3). 
In the \textit{commensurate} phases, the low-energy sectors occupy an enlarged area, where as in the \textit{incommensurate} cases these low-energy sectors become increasingly mixed with high-energy sectors.
To the right of the \textit{incommensurate} relaxation profiles, we show the $\theta_{32}$-dependent fraction of low-energy sectors among all sectors, which increase as $\theta_{32}$ approaches the \textit{commensurate} condition $\theta_{12}=\theta_{32}$. 
}
\label{figS:commensurate}
\end{center}
\end{figure}

For the t3BN systems, we first explore 
the \textit{commensurate} double-moir\'e configurations where the two interfaces share the same moir\'e period.
Since there is no lattice mismatch, the twist angles are the same for both moir\'e interfaces, such that $\theta_{12}=\theta_{32}$.
Similar to the t3G double-moir\'e system~\cite{Leconte_2022, leconte2023commensuration}, the t3BN system has three distinct highly symmetric configurations, $\overline{\text{AAA}}$, $\overline{\text{AAB}}$, and $\overline{\text{AAC}}$, which arise when sliding the top layer with respect to the bottom layer prior to rotation. 
The labels A, B, and C denote the highly symmetric stacking positions obtained by displacing each layer by sliding vectors $\bm{0}$, $\bm{\tau}$, and $2\bm{\tau}$, respectively, measured from the bottom layer, which defines the reference stacking position (A). 
Here, the sliding vector $\bm{\tau} = (0, a_\text{bottom}/\sqrt{3})$ is parallel to the bonding direction of the bottom layer, where $a_\text{bottom}=|\bm{a}_1^{(1)}|$ is the lattice constant of the bottom layer and $\bm{a}_1^{(1)}$ is one of its lattice vectors, as defined in Appendix~\ref{HermannSection}. 
We use an overline on the stacking labels to indicate systems that have the corresponding local stacking as their rotation center, and to distinguish this notation from that without an overline in Fig.~\ref{figS:commensurate}, which is used to define highly symmetric local stackings within a moir\'e pattern.
Aside from the sliding degree of freedom, the t3BN system possesses an additional degree of freedom associated with the polarity of the B and N atoms. 
Depending on the orientations of the hBN layers, this gives rise to three distinct stacking sequences, BN/BN/BN, NB/BN/BN, and BN/NB/BN.
We use a prime ($^{\prime}$) notation for NB alignments, where the corresponding hBN layer is rotated by 60$^{\circ}$ or 180$^{\circ}$. For example, we denote the NB/BN/BN alignments with an AAC rotation center by $\overline{\text{AAC}^{\prime}}$, and the BN/NB/BN configuration with an AAA rotation center by $\overline{\text{AA}^{\prime}\text{A}}$. 

In Fig.~\ref{fig:energy} (a), we show the top-layer sliding-dependent total-energy variation for selected twist angles $\theta_{12}=\theta_{32}=1.08^{\circ},1.54^{\circ}$, and $2.00^{\circ}$. 
The horizontal and vertical axes correspond to the $x$- and $y$-components of the sliding vector, spanning $\left[0,\ a_\text{bottom}\right]$ and $\left[0, \ \sqrt{3}a_\text{bottom}\right]$, respectively, where $a_\text{bottom}=a_\text{BN}$ for t3BN.  
The colormap indicates the total energy per atom ($E_\text{tot}$), and the minimum and maximum values are shown in its colorbar.
All total energies per atom ($E_\text{tot}$) in Fig.~\ref{fig:energy} are globally shifted by a constant value of 6.7151 meV/atom, chosen so that $E_\text{tot}$ for the BN/BN/BN configuration at $\theta_{12}=\theta_{32}=1.08^{\circ}$ is set as zero. 
We observe that the $\overline{\text{AAA}}$, $\overline{\text{AAC}^\prime}$, and $\overline{\text{AA}^\prime\text{A}}$ configurations are the most stable for the BN/BN/BN, NB/BN/BN, and BN/NB/BN systems, respectively, and we note that all higher-energy sliding configurations have a direct barrier-free pathway to these lowest-energy configurations.
We also note that the smaller the twist angle, the more stable the system. 
For the smallest twist angle considered here, $\theta_{12}=\theta_{32}=1.08^{\circ}$, the BN/BN/BN configuration defines the reference energy $E_\text{tot}=0$ by construction, while the NB/BN/BN and BN/NB/BN configurations lie higher by $0.34$ meV/atom and $0.63$ meV/atom, respectively. 
As summarized in Table~\ref{tab:energy}, this behavior is consistent with the reduction of the interface potential energies, which becomes stronger at smaller twist angles. 
For different angle combinations with $\theta_{12}=\theta_{32}$, the energy differences between the most and least stable sliding configurations are nearly the same for BN/BN/BN and BN/NB/BN systems, on the order of $0.1$-$0.3$ meV/atom. 
In constrast, for the NB/BN/BN system, the smallest-angle case $\theta_{12}=\theta_{32}=1.08^{\circ}$ shows a larger sliding-dependent energy variations of $0.5$ meV/atom, which is reduced to $0.37$ and $0.25$ meV/atom for $\theta_{12}=\theta_{32}=1.54^{\circ}$ and $2.00^{\circ}$, respectively.

Above, we focused on the \textit{commensurate} double-moir\'e t3BN systems where $\theta_{12}=\theta_{32}$. However, in general, the moir\'e period induced by the rotation angle between the bottom and middle layers and that induced by the rotation between the middle and top layers are not equal when $\theta_{12}\neq\theta_{32}$, leading to \textit{incommensurate} phases.
Due to the requirement of working with periodic cells in our energy-minimization simulations, we approximate these \text{incommensurate} phases by commensurate approximants, defined by $p$ and $q$ integers satisfying $p L_{12}^{M} = q L_{23}^{M}$. These approximants differ from the true \textit{commensurate} phases with $p=q$, and the commensuration effect was shown to quickly disappear for larger values of $p$ or $q$ in terms of energetics for these commensurate approximants~\cite{Leconte_2022}.
In panel (b) of Fig.~\ref{fig:energy}, we illustrate the total-energy variation as a function of the top-layer rotation angle $\theta_{32}$, while keeping the bottom-layer rotation angle $\theta_{12}$ fixed with respect to the middle layer.
The inset in the top panel schematically defines the rotation angles $\theta_{12}$ and $\theta_{23}$ for the two moir\'e interfaces. 
We explore the BN/BN/BN, NB/BN/BN, and BN/NB/BN configurations from the top to the bottom panels, where the blue, orange, and green curves correspond to fixed bottom twist angles $\theta_{12}=1.08^{\circ}$, $1.54^{\circ}$, and $2.00^{\circ}$, respectively.
For all alignments, the total energies tend to decrease as $\theta_{32}$ approaches zero, which suggests that the system prefers to recover perfect alignment between the top and middle layers. Because of the symmetry of these systems, although not illustrated here, the bottom layer also tends to recover zero-degree alignment. 
As in t3G~\cite{Leconte_2022}, the t3BN system exhibits local energy minima in the double-moir\'e \textit{commensurate} phase, indicated by vertical lines. 
These local minima are accompanied by torque sign reversals, indicating a restoring tendency toward the \textit{commensurate} configuration, while the torque magnitude reflects the strength of the angular stability. 
This suggests that one can, in principle, lock the systems into these \textit{commensurate} phases or even recover the trivially stacked system without any twist between the layers.

\begin{table*}[tbhp]
\label{torque}
\resizebox{0.7\linewidth}{!}{%
\begin{tabular}{c|rrr|rrr|rrr}
\hline
 $\theta_{12}=\theta_{32}$& \multicolumn{3}{|c}{ BN/BN/BN } &  \multicolumn{3}{|c}{ NB/BN/BN} & \multicolumn{3}{|c}{ BN/NB/BN}\\ \hline
 $\theta_{12}\ (^{\circ})$ &   $1.08$     &  $1.53$     & $2.00$ &   $1.08$     &  $1.53$     & $2.00$ &   $1.08$     &  $1.53$     & $2.00$  \\ \hline
$ k_{-}(\theta_{32}) $  
& $-$7.81   &  $-$20.17  &  $-$24.83     
& $-$8.88   &  $-$20.71  &  $-$22.52    
& $-$5.50   &  $-$15.43  &  $-$9.66     \\ \hline
$k_{+}(\theta_{32}) $   
&  93.50    &  75.04    & 58.15     
&  78.07    &  60.53    & 45.75     
&  75.29    &  53.58    & 38.62      \\ \hline
\begin{tabular}[c]{@{}c@{}} $E_b(\theta_{32})$ \\ (meV/atom)\end{tabular}    
& 0.229   &  0.210  &  0.181    
& 0.287   &  0.212  &  0.201     
& 0.164   &  0.121  &  0.119  \\  \hline
 \hline
 $\theta_{12}=\theta_{32}$ & \multicolumn{3}{|c}{ G/BN/G } &  \multicolumn{3}{|c}{ BN/G/NB} & \multicolumn{3}{|c}{ BN/G/BN}\\ \hline
   $\theta_{12}\ (^{\circ})$ &   
   $0.60$     &  $1.27$     & $1.78$ &   
   $0.53$     &  $1.08$     & $1.62$ &   
   $0.53$     &  $1.08$     & $1.62$   \\ \hline
   $ k_{-}(\theta_{32}) $  & 
   $-$1.29 & 11.90  & 5.48 & 
   $-$1.81 &  6.49  & 5.98 &  
   $-$0.26 & 10.22  & 7.99   \\ \hline
   $ k_{+}(\theta_{32}) $  &    
   26.89 & 18.84 & 11.86  & 
   36.72 & 39.59 & 24.29 &  
   35.03 & 35.80 & 21.93 \\ \hline
   \begin{tabular}[c]{@{}c@{}} $E_b(\theta_{32})$ \\ (meV/atom)\end{tabular}    & 
   0.134 & 0.024 & 0.019 & 
   0.175 & 0.096 & 0.057 & 
   0.164 & 0.079 & 0.047 \\ \hline
 \end{tabular}
 }
\resizebox{0.4\linewidth}{!}{
\begin{tabular}{c|rr|rr}
\hline
$\theta_{32}=0.58^{\circ}$ & \multicolumn{2}{|c|}{ G/BN/BN }& \multicolumn{2}{|c|}{ G/BN/NB }  \\ \hline
$\theta_{12}\ (^{\circ})$ &   $+1.20$     &  $-1.20$   &   $+1.20$     &  $-1.20$    \\ \hline
$ k_{-}(\theta_{32}) $ & 
 $-$0.91  & $-$24.39 &
 $-$12.75 & $-$20.27 \\ \hline
$ k_{+}(\theta_{32}) $ &
   24.34 &  34.46 &
   32.88 &  33.58 \\ \hline
\begin{tabular}[c]{@{}c@{}} $E_b(\theta_{32})$ \\ (meV/atom)\end{tabular}  &  
  0.072 & 0.151 &
  0.088 & 0.143 \\ \hline
 \end{tabular}
 }
\caption{
We define the torque constant as $k \equiv {dE_\text{tot}} / { d \theta_{32} }$ and evaluate its values $k_-$ and $k_+$ from the left and right of each local minimum. 
The binding energies $E_b$ at the double-moir\'e \textit{commensurate} configurations are also listed in the last row. 
}
\label{tab:torque}
\end{table*}
In Table.~\ref{tab:torque}, we quantify this angular stability by summarizing the torque constants $k=dE_\text{ tot}/d\theta_{32}$ on both sides of the commensurate angle, together with the binding energy $E_{b}$.
In the case of double-moir\'e {\it commensuration}, a distinct pattern emerges: negative torque constants on the left side and positive torque constants on the right side of the \textit{commensuration} angle, indicating a local minimum in the total energy. 
This behavior is particularly pronounced in t3BN systems, with binding energies ranging from 0.2 to 0.3 meV/atom, comparable to sliding-dependent energy differences.
Here, the binding energies of these {\it commensurate} phases are measured relative to the dashed reference curves, as defined in Sec.~\ref{methodSect}.

To delve into the origin of the local minima in the total energy,
we analyze the relaxation profiles of both double-moir\'e {\it commensurate} and {\it incommensurate} t3BN configurations at $\theta_{12}=1.54^{\circ}$. 
In Fig.~\ref{figS:commensurate}, we divide the local stackings within a double-moir\'e unit cell into 36 regions and represent them using discrete colors, as specified in the figure. 
For each moir\'e interface, the local stacking configurations are grouped into six regions, corresponding to three highly symmetric local stackings (AA, AB, BA) and three intermediate domain walls (DW1, DW2, DW3), 
following the six-region definitions of Ref.~[\onlinecite{PhysRevB.106.115410}].
These relaxation profiles show how the area occupied by low-energy local stackings evolves as the structure moves away from double-moir\'e {\it commensuration}.
As depicted in Fig.~\ref{figS:commensurate}, the contributions from low-energy local stackings are more pronounced in double-moir\'e {\it commensurate} configurations, thanks to the large overlap of the low-energy stacking regions in the stacking maps of the two moir\'e interfaces.  
For instance, in the $\overline{\mathrm{AAA}}$ configuration, AB stacking at the bottom moir\'e interface overlaps with BA stacking at the top moir\'e interface to form ABA local stackings, and similarly AC stacking overlaps with CA stacking to form ACA local stackings, leading to a larger area of ABA and ACA low-energy local stackings. 
Similarly, the $\overline{\mathrm{AAC}'}$ configuration has a larger contribution from ABC$'$ and ACC$'$ stackings, while the $\overline{\mathrm{AA}'\mathrm{A}}$ configuration is characterized by a larger area of AA$'$A and AB$'$A stackings. 
Conversely, the double-moir\'e {\it incommensurate} systems display more intricate patterns of local stackings due to the supermoir\'e structure, which reduces the overlap of low-energy stacking regions between the two moir\'e interfaces and thus diminishes their overall contribution.
To the right of the {\it incommensurate} local stacking maps, we present the $\theta_{32}$-dependent ratio of low-energy stacking areas, which decreases as the top twist angle $\theta_{32}$ deviates from the bottom twist angle $\theta_{12}$.
This trend explains why the total energy is minimized at the double-moir\'e {\it commensurate} configuration and, equivalently, why deviations from this condition generate a restoring torque toward the \textit{commensurate} configuration.
Since hBN is a polar material, one may expect electrostatic effects to influence, qualitatively or quantitatively, the observed local energy minima. 
In Appendix \ref{electrostatic}, we summarize benchmark calculations with pairwise Coulomb interaction potentials, focusing on the $\overline{\text{AAA}}$ configuration of t3BN.
Specifically, we consider two pairwise potentials implemented in LAMMPS. 
The first is the \texttt{coul/shield} pair style, which describes the short-range Coulomb interactions with a cutoff distance $r_\text{cut}$\cite{coul_shield_Leven1, coul_shield_Leven2, coul_shield_Maaravi} and a shielding parameter $\lambda$\cite{coul_shield_Maaravi}. 
The second is the \texttt{coul/long} pair style for long-range Coulomb interactions, in which the contributions for $r_{ij}>r_\text{cut}$ are evaluated in reciprocal space using a particle-particle particle-mesh (\texttt{pppm}) solver\cite{coul_long_pppm_Hockney}, enabling an efficient fast Fourier transformation implementation\cite{coul_long_pppm_Pollock}.   
We find that the impact of these electrostatic effects is mostly quantitative. The local energy minima at the double-moir\'e \textit{commensuration} persist for both short- and long-range Coulomb interaction potentials and become even deeper for the long-range case. 
Applying an external electric field (up to $\sim 0.15$ V/nm) also preserves these energy dips, although those results are not shown explicitly here.
This suggests that double-moir\'e \textit{commensuration} can be further stabilized in polar materials. 
Interestingly, we find that the long-range Coulomb interaction modifies the corrugation profiles of the relaxed double-moir\'e \textit{commensurate} configurations.
These corrugation profiles separate the stable systems into two families, namely the breathing mode and the bending mode, 
which were previously shown to impact the energetics of the system~\cite{bending_corrugation}.
In the absence of electrostatic effects, the breathing-mode-like structure, which is mirror symmetric, is relatively less stable than the bending-mode-like structure, in which the mirror symmetry is broken. 
See Table.~\ref{tab:energy}, where the energies for the mirror-symmetry-broken (MSB) cases are indicated by an asterisk(*) in their stacking labels. 
This ordering remains in the presence of the short-range Coulomb interactions.
However, when the long-range Coulomb interactions are included, the  breathing-mode-like structure becomes more energetically stable than the bending-mode-like configuration. 
\subsection{Unequal lattice constant heterolayer systems}
\label{sec:BNandG}

\begin{figure*}[tbhp]
\begin{center}
\includegraphics[width=0.8\textwidth]{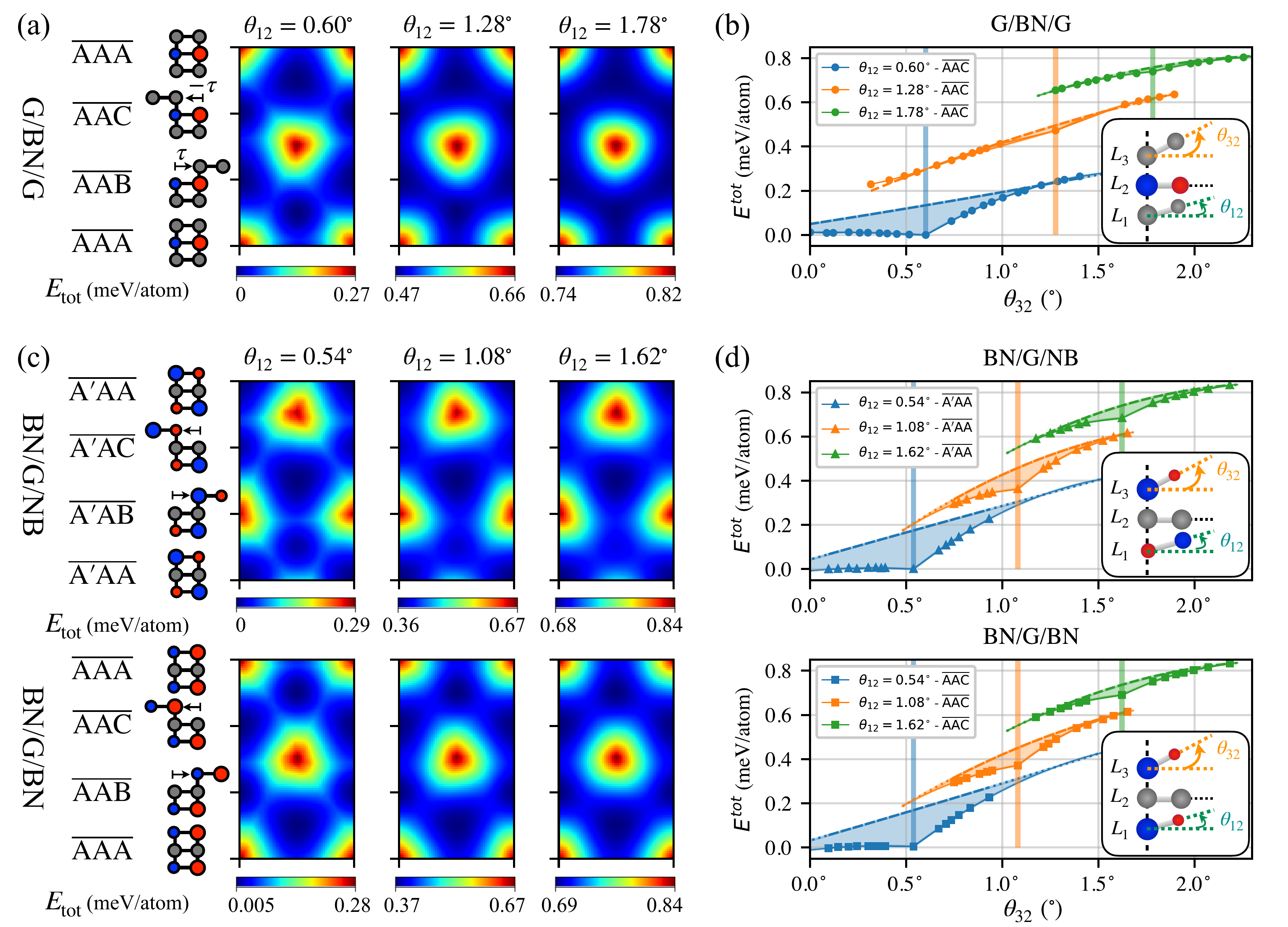}
\caption{(color online)
Total energies per atom for G/BN/G, BN/G/NB, and BN/G/BN systems as a function of sliding and top twist angle $\theta_{32}$ with respect to the middle layer. 
(a) Sliding-dependent energy maps for \textit{commensurate} G/BN/G structures at $\theta_{12}=\theta_{23}=0.6,\ 1.28,\ 1.78^{\circ}$ show that the $\overline{\text{AAC}}$ configuration are the most stable and smaller twist angles yield lower total energies. 
(b) $\theta_{32}$-dependent transitions from \textit{incommensurate} to \textit{commensurate} G/BN/G structures, showing the local energy minima at the \textit{commensurate} conditions (indicated by vertical lines). 
The curves to the left and right of each \textit{commensurate} angle are extrapolated using parabolic fits, and we approximate reference total-energy curves for the \textit{incommensurate} regime from the parabolic fits at the left and right edges. 
The binding energies, defined by the energy differences between these smooth reference curves and the actual total energies at the \textit{commensuration}, are nearly zero in the larger-angle regime ($\theta_{12}> 1^{\circ}$) but become sizable at the small angle $\theta_{12}\approx 0.6^{\circ}$. 
(c), (d) Same as panels (a), (b), respectively, but for the BN/G/NB and BN/G/BN structures at selected twist angles $\theta_{12}=\theta_{23}=0.54,\ 1.08,\ 1.62^{\circ}$. 
The $\overline{\text{A}^{\prime}\text{AA}}$ and $\overline{\text{AAC}}$ configurations for BN/G/NB and BN/G/BN, respectively, are energetically more favorable than the other configurations, with BN/G/NB being slightly more stable than than BN/G/BN by about 0.005 meV/atom.
The large binding energy at the small \textit{commensuration} angle $\theta_{12}=0.54^\circ$) decreases as the \textit{commensuration} angle increases, similarly to the G/BN/G case.
The total energies per atom are globally shifted by a constant value of 7.1931 eV/atom for the panels (a) and (b), and 6.9588 eV/atom for panels (c) and (d), so that the energy minimum shown in each panel is set to zero. 
}
\label{fig:energy_GBNG_BNGBN}
\end{center}
\end{figure*}

\begin{figure*}[tbhp]
\begin{center}
\includegraphics[width=0.8\textwidth]{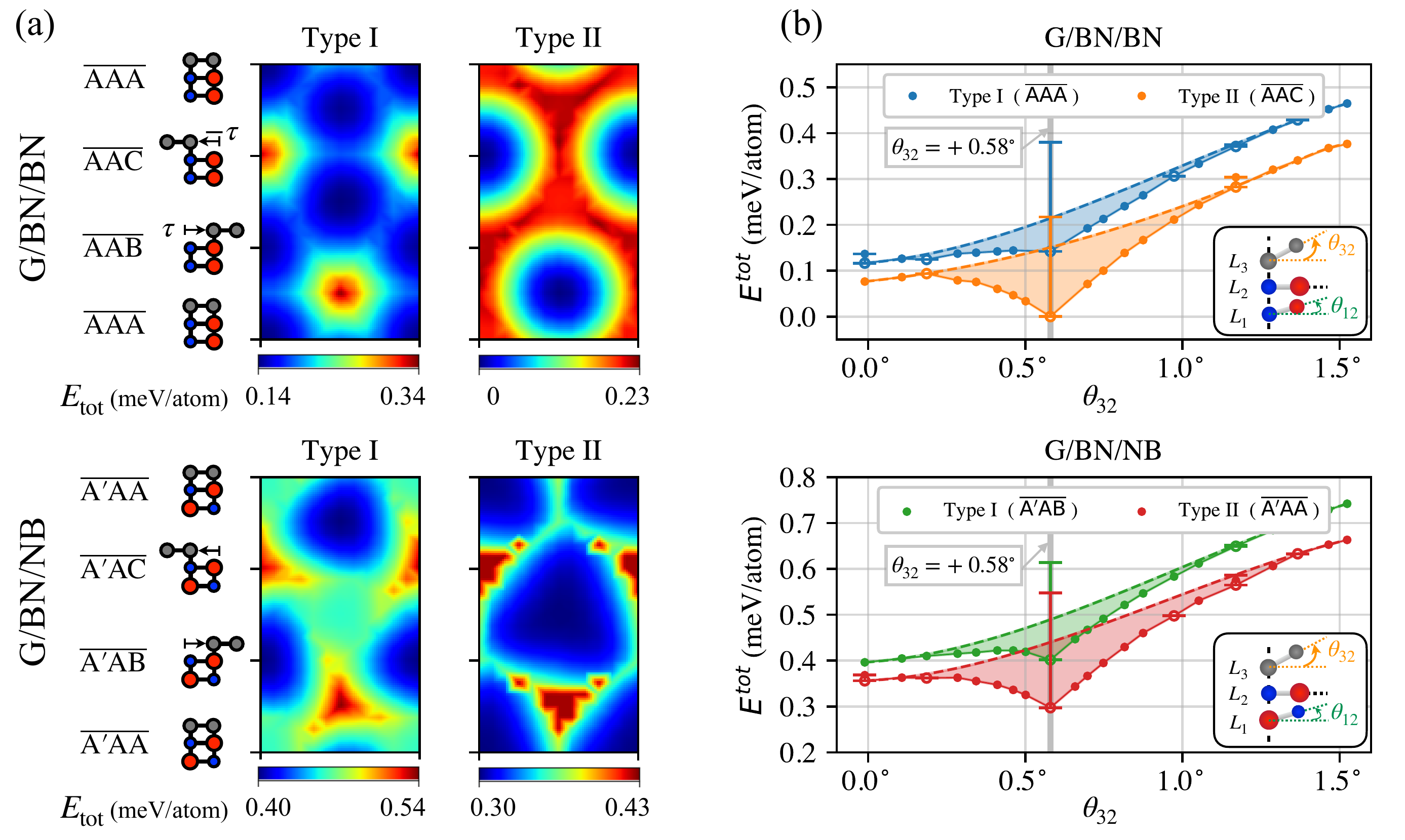}
\caption{(color online)
Total energies per atom for G/BN/BN and G/BN/NB systems as a function of (a) sliding and (b) the top-layer twist angle $\theta_{32}$ with respect to the middle layer.
Each system is considered for two twist-angle combinations with bottom twist angles $\theta_{12} = \pm 1.20^{\circ}$ (type I and type II) and a common top twist angle $\theta_{32} = 0.58^{\circ}$. 
All four cases exhibit local energy minima at the double-moir\'e \textit{commensurate} condition (indicated by vertical gray lines). 
For both stacking sequences, type II is energetically more favorable than type I, with the energy differences on the order of $0.1$ meV/atom, and G/BN/BN is more stable than G/BN/NB by about $0.3$ meV/atom. 
Unlike in the other systems, relatively small \textit{incommensurate} simulation cells, with lattice constants two or three times larger than that of the \textit{commensurate} cell, produce non-smooth trends, and their sliding-dependent energy barriers remain finite. 
This occurs because the two moir\'e lengths are very similar, with $p\approx q$ in the relation $p L_{12}^{M} = q L_{23}^{M}$.
Open and bar markers connected with vertical lines indicate, respectively, the minimum and maximum sliding-dependent total energies for these smaller \textit{incommensurate} simulation cells, 
while filled markers denote the total energies per atom for the lowest-energy sliding configurations of the \textit{commensurate} cell.
The total energies per atom are globally shifted by a constant value of 6.9573 eV/atom so that the minimum energy among these cases is set to zero.
}
\label{fig:energy_GBNBN_GBNNB}
\end{center}
\end{figure*}

For the lattice-mismatched systems, we first focus on G/BN/G, BN/G/NB, and BN/G/BN where the two moir\'es are generated by the same combination of G and hBN, \textit{i.e.}, a mirror-symmetric stacking of layers with respect to the middle layer. 
Owing to this symmetry, the \textit{commensurate} phase can be obtained simply by setting $\theta_{12}=\theta_{32}$. 
In panels (a) and (c) of Fig.~\ref{fig:energy_GBNG_BNGBN}, we show sliding maps for these systems, analogous to those in Fig.~\ref{fig:energy} for t3BN. 
We find that the $\overline{\text{AAC}}$, $\overline{\text{A}^\prime\text{AA}}$ and $\overline{\text{AAC}}$ configurations are the most stable sliding configurations for G/BN/G, BN/G/NB, and BN/G/BN, respectively. 
For the \textit{commensurate} phases of these systems, 
as in t3BN, the smaller the angle, the lower the total energies.
The sliding-dependent energy variations are on the order of $\sim 0.3$ meV/atom for the smaller twist angle around $\theta_{12}=\theta_{32}\approx 0.6^{\circ}$, and they are reduced to about $\sim0.1$ meV/atom for the larger-angle regime with $\theta_{12}=\theta_{32}>1.6^{\circ}$.  
In panels (b) and (d), we illustrate how the total energy evolves in the \textit{incommensurate} regime as a function of $\theta_{32}$. 
We generally observe, as in t3BN, that the local minima appear near the double-moir\'e \textit{commensurate} angles, while the aligned configuration remains the globally most stable one. 
However, this trend is weakened for smaller values of $\theta_{12}\le0.6^{\circ}$, where the total energy at the double-moir\'e \textit{commensuration} remains comparable to that of perfect alignment.
While the energy differences become very small in this regime, the commensurate configuration in G/BN/G still exhibits the lowest energy within the range of twist angles considered here.
This suggests a possible scenario in which such energetic stabilization away from alignment in a three-layer system could be leveraged to realize stable bilayer systems with twist angles away from alignment. 
We note that our calculations are performed with periodic boundary conditions, which are distinct from the metastable phase at an approximate twist angle of $0.6^{\circ}$ observed in a graphene flake on an hBN substrate~\cite{woods2016,jharapla2025geometriccontrolmoiretwist}.
Given that the freestanding G/BN bilayer has its total-energy minimum near zero twist angle~\cite{GBN}, 
the observed global minimum at $\theta_{12}\le0.6^{\circ}$ can be interpreted as arising from a double-moir\'e \textit{commensuration} effect.
This interpretation is further supported by the comparison of relaxation effects in G/BN bilayers presented in Appendix~\ref{strain}.
For larger angles, the local dips become progressively less pronounced, 
as highlighted by the vanishing area between the dashed-line reference curve and the actual data points, in particular for the G/BN/G system.
This indicates that the top layer may no longer lock into \textit{commensuration}, 
as the torque constants on the left and right sides of the dip have the same sign and the binding energies are reduced to (almost) zero.
The binding energies for these cases are summarized in Table~\ref{tab:torque}.
Calculating additional \textit{incommensurate} angles closer to the \textit{commensuration} point to confirm this assertion lies beyond our current computational resources, due to the size of commensurate approximants required. 
We then finally switch our attention to the G and hBN stacking combinations for which the \textit{commensurate} phase occurs for $\theta_{12} \neq \theta_{32}$ due to the asymmetry between the bottom and top moir\'es, \textit{i.e.}, G/BN/BN and G/BN/NB. 
We consider four configurations, corresponding to two orientations of the bottom hBN layer and two choices of rotational directions.
All four configurations similarly exhibit a local minimum of the total energy at the double-moir\'e \textit{commensurate} angle $\theta_{32}=0.58^{\circ}$ for both type I and type II with $\theta_{12}=\pm 1.20^{\circ}$, as shown in Fig.~\ref{fig:energy_GBNBN_GBNNB}.
In addition, the global energy minimum occur at the double-moir\'e \textit{commensuration} rather than at perfect alignment, 
in line with the energetic preference for the double-moir\'e \textit{commensuration} observed in the symmetric G/BN/G, BN/G/NB, and BN/G/BN cases discussed in Fig.~\ref{fig:energy_GBNG_BNGBN}. 
The corresponding binding energies remain comparable, around 0.1-0.15 meV/atom.
We observe the onset of a decreasing energy trend for $\theta_{32}>1.5^\circ$, but we have not investigated this regime further because it is difficult to construct commensurate cells.
This difficulty arises because we keep $\theta_{12}$ fixed and impose a maximum strain of 0.03\% on the lattice constant $|\bm{a}_1^{(3)}|$ of the top layer. 
In these cases, however, the energy profiles for \textit{incommensurate} angles do not form a smooth total-energy curve when we follow the most stable sliding configures of the \textit{commensurate} phases, as illustrated with the filled markers.
For example, the total-energy data points around $\theta_{32}\approx -0.01^{\circ},\ 0.18^{\circ},\ 1.17^{\circ}$ noticeably deviate from an otherwise smooth trend. 
These data points correspond to the relatively small \textit{incommensurate} cells, whose lattice constants are $p(\le3)$- and $q(\le3)$-times the bottom and top moir\'e lattice constants, respectively, when $p L_{12}^{M} = q L_{23}^{M}$. 
Unlike the other systems, where sliding effects are less prominent in the \textit{incommensurate} regime, the G/BN/BN and G/BN/NB configurations at these smaller \textit{incommensurate} supercells exhibit a relatively large energy barrier that depends on the stacking configuration at the rotation center.
For the $p\le 3$ cases, we examine highly symmetric sliding configurations, and the minimum and maximum total energies per atom with respect to the sliding are indicated by open and bar markers, respectively, in the panel (b).
The torque coefficients for the lattice-mismatched systems are also reported in Table~\ref{tab:torque}. 
In contrast to the lattice-matched systems, including t3BN, the opposite signs of the torque coefficients persist over a broader angular range in the lattice-mismatched systems, indicating enhanced angular stability around the double-moir\'e \textit{commensurate} configuration.
This extended restoring tendency is consistent with the fact that the \textit{commensurate} local minima become global minima in these systems, as clearly seen in the $0.54^\circ$ curves in Fig.~\ref{fig:energy_GBNG_BNGBN} (d) for the BN/G/BN and BN/G/NB systems, and in the $0.58^\circ$ curves in Fig.~\ref{fig:energy_GBNBN_GBNNB} (b) for the G/BN/BN and G/BN/NB systems.

\section{Summary and Conclusion}
\label{sec:conclusions}
We have shown that combining different moir\'e patterns built from graphene and hBN can lead to robust angle locking in double-moir\'e \textit{commensurate} phases. In t3BN, where no intrinsic lattice mismatch is present, double-moir\'e \textit{commensurate} configurations with $\theta_{12} = \theta_{32}$ exhibit binding energies on the order of $\sim 0.2$ meV/atom over a broad twist-angle range, indicating that such states should be experimentally accessible.
In the symmetric lattice-mismatched systems G/BN/G, BN/G/BN, and BN/G/NB, where \textit{commensuration} still occurs at $\theta_{12} = \theta_{32}$, local energy minima are pronouncedly found only for small twist angles $\theta_{12} = \theta_{32} \simeq 0.6^{\circ}$, with comparable binding energies of about $0.1$–$0.2$ meV/atom. 
In the asymmetric lattice-mismatched systems G/BN/BN and G/BN/NB, where \textit{commensuration} occurs at $\theta_{12} \neq \theta_{32}$, the angle-locking effects at the \textit{commensurate} angles persist, with binding energies of about 0.1–0.15 meV/atom, sufficiently large to lock these systems into their double-moir\'e \textit{commensurate} phases.
Benchmark calculations for the t3BN-$\overline{\text{AAA}}$ configuration indicate that these angle-locking effects persist in the presence of both short- and long-range Coulomb interactions, modeled in LAMMPS via the \texttt{coul/shield} and \texttt{coul/long} pair styles, suggesting that electrostatics in polar stacks may enhance 
stabilization at double-moir\'e \textit{commensuration}.

To clarify the microscopic origin of these local energy minima, we analyzed the relaxation profiles in both \textit{commensurate} and \textit{incommensurate} phases. 
In the double-moir\'e \textit{commensurate} case, the low-energy local stacking regions at the two moir\'e interfaces substantially overlap, increasing the weight of energetically favorable domains and thereby lowering the total energy. This mechanism selects specific sliding configurations as the most stable, such as the $\overline{\text{AAA}}$ stacking in BN/BN/BN, which benefits from enhanced contributions of ABA and ACA local stackings. 
In \textit{incommensurate} phases, the overlap of low-energy stackings between the two interfaces is diminished, the weights of individual domains become redistributed more uniformly, and the energy gains from relaxation are effectively averaged away. 
Consequently, the total energy in the \textit{incommensurate} regime is relatively insensitive to sliding, pointing to enhanced superlubricity, consistent with previous conclusions for t3G.
Although a fully systematic quantification for all systems studied here is computationally demanding, our results generally indicate that, when comparing the sliding-dependent energy scales to the rotational binding energies, the latter dominate in determining the most stable configurations.
However, depending on the specific system, smaller \textit{incommensurate} simulation cells can still exhibit non-negligible sliding-dependent energy barriers. 
This is the case, for example, for the \textit{incommensurate} supercells of the  G/BN/BN and G/BN/NB systems with $p \le 3$ and $q \approx p$ satisfying $p L_{12}^{M} = q L_{23}^{M}$.
These results therefore support the view that the energy locking at double-moir\'e commensuration is generally expected in double-moir\'e trilayers, while also suggesting that, in some systems, sliding-dependent energy barriers in small \textit{incommensurate} cells can become a significant contribution.  

We finally note that the location of the global total-energy minimum is highly system-specific.
In t3BN, the global energy minimum occurs when one interface returns to perfect crystallographic alignment, while double-moir\'e \textit{commensurate} angles only produce local minima, in line with the predictions for t3G~\cite{leconte2023commensuration}, one of the lattice-matched systems.
By contrast, the lattice-mismatched G/hBN trilayers considered here show the energetic preference for the double-moir\'e \textit{commensuration}.
For example, the small-angle double-moir\'e \textit{commensurate} phase near $\theta_{12}=\theta_{32}\approx0.6^\circ$ exhibits a comparable or slightly lower energy than the perfectly aligned configuration in the symmetric lattice-mismatched systems G/BN/G, BN/G/BN, and BN/G/NB.
This tendency becomes even more pronounced in the asymmetric lattice-mismatched systems G/BN/BN and G/BN/NB, where the global energy minimum occurs at \textit{commensurate} angles around $\theta_{32}\approx 0.6^{\circ}$ for both type I and type II, with $\theta_{12}=\pm 1.20^{\circ}$.
The behavior at G/BN interfaces near a twist angle of $\sim0.6^{\circ}$ therefore deserves particular attention, since metastable states near this angle have been reported for a graphene flake on hBN substrates~\cite{woods2016, jharapla2025geometriccontrolmoiretwist} and for a relaxed graphene layer on a rigid hBN layer~\cite{GBN}.
However, the global energy minima in our lattice-mismatched trilayers stem from double-moir\'e \textit{commensuration} effects.
Because our calculations employ periodic simulation cells, we can exclude the metastable configurations specific to a finite graphene flake on an hBN substrate, which are known to arise from coincident orientations between moir\'e lattices and flake edges.
For a relaxed graphene layer on a rigid hBN layer, the energy minimum at $0.6^{\circ}$ relative to $0^{\circ}$ is about 0.015 meV/atom, roughly an order of magnitude smaller than the binding energies $\sim0.15$ meV/atom observed at the \textit{commensuration}.
This suggests that the previously reported rigid-substrate mechanism is unlikely to account for our global minima.
Moreover, in the previous t2G/BN study~\cite{leconte2023commensuration}, the \textit{commensurate} angle $\theta_{12}\simeq \pm 0.6^{\circ}$ at the G/hBN interface did not correspond to a global energy minimum, even though the associated binding energies are comparable to those in the G/BN/BN and G/BN/NB configurations.  

Taken together, our results on G and hBN stacking combinations indicate that many other double-moir\'e systems are likely to display similar angle-locking and \textit{commensuration}-driven energetics. 
At the same time, the stability of a given \textit{commensurate} angle combination is strongly system-dependent.
Some configurations may be readily realizable, whereas others may be challenging, and in certain cases the double-moir\'e \textit{commensurate} state may even be stabilized as the global minimum.
Therefore, targeted calculations of torque and binding energies can provide valuable guidance in determining whether a desired double-moir\'e configuration is likely to be robust, angle-locked, and experimentally realizable.

\begin{acknowledgments}
This work was supported by the Korean NRF through the Grant
No. NRF-2020R1A5A1016518 and RS-2026-25474048.
We acknowledge computational support from KISTI Grant No. KSC-2021-CRE-0389 and by the resources of Urban Big data and AI Institute (UBAI) at UOS. J.J also acknowledges support by the Korean Ministry of Land, Infrastructure and Transport(MOLIT) from the Innovative Talent Education Program for Smart Cities.
\end{acknowledgments}

\section*{Data Availability}
The data and scripts that support the findings of this article are openly available~\cite{data_repo}.

\appendix
\renewcommand\thefigure{S.\arabic{figure}}    

\setcounter{figure}{0}

\newpage
\section{Commensurate simulation cells}
\label{HermannSection}

%


The commensurate simulation cells satisfy the relation~\cite{Hermann:2012dy, PhysRevB.106.115410}
\begin{equation}
    \begin{aligned}
        \begin{pmatrix}\bm{R}_1 \\ \bm{R}_2\end{pmatrix} 
        &= \begin{pmatrix} i  & j \\ -j & i+j \end{pmatrix} \begin{pmatrix}\bm{a}_1^{(1)} \\ \bm{a}_2^{(1)}\end{pmatrix} \\
        &= \begin{pmatrix} i^{\prime}  & j^{\prime} \\ -j^{\prime} & i^{\prime}+j^{\prime} \end{pmatrix} \begin{pmatrix}\bm{a}_1^{(2)} \\ \bm{a}_2^{(2)}\end{pmatrix} \\
        &= \begin{pmatrix} i^{\prime\prime}  & j^{\prime\prime} \\ -j^{\prime\prime} & i^{\prime\prime}+j^{\prime\prime} \end{pmatrix} \begin{pmatrix}\bm{a}_1^{(3)} \\ \bm{a}_2^{(3)}\end{pmatrix} 
    \end{aligned}
    \label{eq:condition}
\end{equation}
where $\bm{R}_i$ with $i=1,2$ denote the lattice vectors of the commensurate cell, and $\bm{a}_i^{(\ell)}$ with $i=1,2$ for the lattice vectors of the unit lattice vectors of the layer $\ell$, which are rotated by $\theta_{\ell}$.
We set the middle layer as the reference frame to measure the lattice mismatch of $\alpha_{12}$ and $\alpha_{32}$, and the twist angle of $\theta_{12}$ and $\theta_{32}$ for the two moir\'e interfaces, one between the bottom and middle layers and the other between the top and middle layers.
Fixing the lattice constants of the middle layer $|\bm{a}_1^{(2)}|$ as $a_G=2.46019~\AA$ (graphene) or $a_{BN}=2.505~\AA$ (hBN), we calculate the lattice mismatch and the twist angles for the bottom and top layers with respect to the middle layer as
\begin{equation}
    \begin{aligned}
        \alpha_{12}&=\frac{|\bm{a}_1^{(1)}|}{|\bm{a}_1^{(2)}|} = \sqrt{\frac{{i^{\prime}}^2+{j^{\prime}}^2+i^{\prime} j^{\prime}}{i^2+j^2+i j}} , \\
        \alpha_{32}&=\frac{|\bm{a}_1^{(3)}|}{|\bm{a}_1^{(2)}|} = \sqrt{\frac{{i^{\prime}}^2+{j^{\prime}}^2+i^{\prime} j^{\prime}}{{i^{\prime\prime}}^2+{j^{\prime\prime}}^2+{i^{\prime\prime}} {j^{\prime\prime}}}} ,\\
        \theta_{12} &= \theta_{1}-\theta_{2} = \cos^{-1}\left[\frac{ 2 i i^{\prime} + 2 j j^{\prime} + i j^{\prime} + j i^{\prime} }{2 \alpha_{12}(i^2+j^2+i j)}\right],\\
        \theta_{32} &= \theta_{3}-\theta_{2} = \cos^{-1}\left[\frac{ 2 i^{\prime\prime} i^{\prime} + 2 j^{\prime\prime} j^{\prime} + i^{\prime\prime} j^{\prime} + j^{\prime\prime} i^{\prime} }{2 \alpha_{32}({i^{\prime\prime}}^2+{j^{\prime\prime}}^2+{i^{\prime\prime}}{j^{\prime\prime}})}\right],\\
    \end{aligned}
     \label{eq:lattice}
\end{equation}
In Table~\ref{HermannIndices1}, and \ref{HermannIndices2}, we summarize the indices for each of our commensurate cells.

\begin{table*}[tbhp]
\begin{minipage}{.48\linewidth}
\resizebox{1.0\columnwidth}{!}{%
\begin{tabular}{|lccccc|}
\multicolumn{6}{c}{t3BN} \\
\hline
$\theta_{12}$ ($^{\circ}$) &$\theta_{32}$ ($^{\circ}$) & ($i$ $j$ $i'$ $j'$ $i''$ $j''$) & $|\bm{a}_1^{(3)}|$($\AA$) & \# atoms  & $\lambda$\\\hline
1.084549 & 0.271131 & 124 120 120 124 121 123  & 2.505843   &   267930   &   4$\lambda_1$    \\ 
         & 0.361512 & 93 90 90 93 91 92        & 2.505859   &   150710   &   3$\lambda_1$    \\ 
         & 0.542275 & 62 60 60 62 61 61        & 2.505871   &   66982   &   2$\lambda_1$    \\ 
         & 0.723037 & 93 90 90 93 92 91        & 2.505859   &   150710   &   3$\lambda_1$    \\ 
         & 0.813418 & 124 120 120 124 123 121  & 2.505843   &   267930   &   4$\lambda_1$    \\ 
         & 0.976099 & 310 300 300 310 309 301  & 2.505799   &   1674582   &   10$\lambda_1$    \\ 
         & {\bf 1.084549} & {\bf 31 30 30 31 31 30 } & {\bf 2.505759}   &   {\bf 16746 }   & {\bf 1}$\lambda_1$   \\ 
         & 1.192995 & 310 300 300 310 311 299  & 2.505710   &   1674622   &   10$\lambda_1$    \\ 
         & 1.446017 & 93 90 90 93 94 89        & 2.505560   &   150722   &   3$\lambda_1$    \\ 
         & 1.626726 & 62 60 60 62 63 59        & 2.505422   &   66990   &   2$\lambda_1$    \\ 
         & 1.807414 & 93 90 90 93 95 88        & 2.505260   &   150734   &   3$\lambda_1$    \\ 
         & 2.168710 & 31 30 30 31 32 29        & 2.504862   &   16750   &   1$\lambda_1$    \\ 
         & 2.710402 & 62 60 60 62 65 57        & 2.504077   &   67014   &   2$\lambda_1$    \\ 
 \hline 
 1.538500 & 0.384608 & 88 84 84 88 85 87        & 2.505928   &   133146   &   4$\lambda_2$    \\ 
          & 0.512820 & 66 63 63 66 64 65        & 2.505960   &   74894   &   3$\lambda_2$    \\ 
          & 0.769250 & 44 42 42 44 43 43        & 2.505985   &   33286   &   2$\lambda_2$    \\ 
          & 1.025680 & 66 63 63 66 65 64        & 2.505960   &   74894   &   3$\lambda_2$    \\ 
          & 1.230818 & 110 105 105 110 109 106  & 2.505904   &   208042   &   5$\lambda_2$    \\ 
          & 1.435943 & 330 315 315 330 329 316  & 2.505815   &   1872422   &   15$\lambda_2$    \\ 
          & {\bf 1.538500 }& {\bf 22 21 21 22 22 21 } & {\bf 2.505759 }  &   {\bf 8322}   &   {\bf 1}$\lambda_2$    \\ 
          & 1.641052 & 330 315 315 330 331 314  & 2.505695   &   1872482   &   15$\lambda_2$    \\ 
          & 1.794869 & 132 126 126 132 133 125  & 2.505583   &   299606   &   6$\lambda_2$    \\ 
          & 1.846138 & 110 105 105 110 111 104  & 2.505542   &   208062   &   5$\lambda_2$    \\ 
          & 1.923038 & 88 84 84 88 89 83        & 2.505477   &   133162   &   4$\lambda_2$    \\ 
          & 2.307473 & 44 42 42 44 45 41        & 2.505082   &   33294   &   2$\lambda_2$    \\ \hline 
2.004628 & 0.668179 & 51 48 48 51 49 50         & 2.506100   &   44114   &   3$\lambda_3$    \\ 
         & 1.002314 & 34 32 32 34 33 33         & 2.506142   &   19606   &   2$\lambda_3$    \\ 
         & 1.336449 & 51 48 48 51 50 49         & 2.506100   &   44114   &   3$\lambda_3$    \\ 
         & 1.754083 & 136 128 128 136 135 129   & 2.505927   &   313714   &   8$\lambda_3$    \\ 
         & 1.904414 & 340 320 320 340 339 321   & 2.505832   &   1960762   &   20$\lambda_3$    \\ 
         & {\bf 2.004628} & {\bf 17 16 16 17 17 16 } & {\bf 2.505759}   &   {\bf 4902}   &  {\bf  1}$\lambda_3$    \\ 
         & 2.104836 & 340 320 320 340 341 319   & 2.505678   &   1960842   &   20$\lambda_3$    \\ 
         & 2.255134 & 136 128 128 136 137 127   & 2.505543   &   313746   &   8$\lambda_3$    \\ 
         & 2.338626 & 102 96 96 102 103 95      & 2.505461   &   176486   &   6$\lambda_3$    \\ 
         & 2.505593 & 68 64 64 68 69 63         & 2.505280   &   78442   &   4$\lambda_3$    \\ 
         & 2.672534 & 51 48 48 51 52 47         & 2.505078   &   44126   &   3$\lambda_3$    \\ 
         & 3.006329 & 34 32 32 34 35 31         & 2.504610   &   19614   &   2$\lambda_3$    \\ \hline
\end{tabular}
}
\end{minipage}
\begin{minipage}{.48\linewidth}
\resizebox{1.0\columnwidth}{!}{%
\begin{tabular}{|lccccc|}
\multicolumn{6}{c}{BN/G/BN} \\
\hline
$\theta_{12}$ ($^{\circ}$) &$\theta_{32}$ ($^{\circ}$) & ($i$ $j$ $i^{\prime}$ $j^{\prime}$ $i^{\prime\prime}$ $j^{\prime\prime}$) & $|\bm{a}_1^{(3)}|$($\AA$) & \# atoms  & $\lambda$\\\hline
 0.537754 & 0.095624 & 208  240  208 248  205 243  & 2.504330   &   614272   &   8$\lambda_4$    \\ 
       & 0.144734 & 78  90  78 93  77 91  & 2.504434   &   86382   &   3$\lambda_4$    \\ 
       & 0.200864 & 182  210  182 217  180 212  & 2.504551   &   470302   &   7$\lambda_4$    \\ 
       & 0.242966 & 104  120  104 124  103 121  & 2.504637   &   153568   &   4$\lambda_4$    \\ 
       & 0.301913 & 130  150  130 155  129 151  & 2.504755   &   239950   &   5$\lambda_4$    \\ 
       & 0.341214 & 156  180  156 186  155 181  & 2.504833   &   345528   &   6$\lambda_4$    \\ 
       & 0.369287 & 182  210  182 217  181 211  & 2.504887   &   470302   &   7$\lambda_4$    \\ 
       & 0.390343 & 208  240  208 248  207 241  & 2.504928   &   614272   &   8$\lambda_4$    \\ 
       & {\bf 0.537754} & {\bf 26  30  26 31  26 30}  & {\bf 2.505202}   &  {\bf  14310}   &  {\bf  1}$\lambda_4$    \\ 
       & 0.668812 & 234  270  234 279  235 269  & 2.505432   &   777438   &   9$\lambda_4$    \\ 
       & 0.706262 & 182  210  182 217  183 209  & 2.505495   &   470302   &   7$\lambda_4$    \\ 
       & 0.734350 & 156 180 156 186  157 179  & 2.505542   &   515114   &   6$\lambda_4$    \\ 
       & 0.773675 & 130  150  130 155  131 149  & 2.505606   &   239950   &   5$\lambda_4$    \\ 
       & 0.832667 & 104  120  104 124  105 119  & 2.505701   &   153568   &   4$\lambda_4$    \\ 
       & 0.930997 & 78  90  78 93  79 89  & 2.505852   &   128768   &   3$\lambda_4$    \\
 \hline 
 1.080027 & 0.747935 & 384  120  384 132  382 123  & 2.504312   &   847008   &   12$\lambda_5$    \\ 
       & 0.773475 & 416  130  416 143  414 133  & 2.504363   &   994058   &   13$\lambda_5$    \\ 
       & 0.880752 & 640  200  640 220  638 203  & 2.504570   &   2352800   &   20$\lambda_5$    \\ 
       & 0.920602 & 800  250  800 275  798 253  & 2.504645   &   3676250   &   25$\lambda_5$    \\ 
       & 0.947170 & 960  300  960 330  958 303  & 2.504694   &   5293800   &   30$\lambda_5$    \\ 
       & 0.814340 & 480  150  480 165  478 153  & 2.504443   &   1323450   &   15$\lambda_5$    \\ 
       & {\bf 1.080027} & {\bf 32  10  32 11  32 10 } & {\bf 2.504932 }  &   {\bf 8770 }  &   {\bf 1}$\lambda_5$    \\ 
       & 1.212907 & 960  300  960 330  962 297  & 2.505157   &   5293800   &   30$\lambda_5$    \\ 
       & 1.239486 & 800  250  800 275  802 247  & 2.505200   &   3676250   &   25$\lambda_5$    \\ 
       & 1.279357 & 640  200  640 220  642 197  & 2.505264   &   2352800   &   20$\lambda_5$    \\ 
       & 1.386709 & 416  130  416 143  418 127  & 2.505430   &   994058   &   13$\lambda_5$    \\ 
       & 1.442482 & 352  110  352 121  354 107  & 2.505513   &   711722   &   11$\lambda_5$    \\ 
       & 1.523050 & 288  90  288 99  290 87  & 2.505628   &   476442   &   9$\lambda_5$    \\ 
       & 1.578444 & 256  80  256 88  258 77  & 2.505705   &   376448   &   8$\lambda_5$    \\ 
       & 1.649671 & 224  70  224 77  226 67  & 2.505800   &   288218   &   7$\lambda_5$    \\      
 \hline 
 1.622768 & 1.174727 & 140  155  135 165  138 157  & 2.504324   &   266100   &   5$\lambda_6$    \\ 
       & 1.249384 & 84  93  81 99  83 94  & 2.504443   &   95796   &   3$\lambda_6$    \\ 
       & 1.302715 & 196  217  189 231  194 219  & 2.504525   &   521556   &   7$\lambda_6$    \\ 
       & 1.342715 & 112  124  108 132  111 125  & 2.504586   &   170304   &   4$\lambda_6$    \\ 
       & 1.398718 & 140  155  135 165  139 156  & 2.504669   &   266100   &   5$\lambda_6$    \\ 
       & 1.436056 & 168  186  162 198  167 187  & 2.504722   &   383184   &   6$\lambda_6$    \\ 
       &{\bf 1.622768}&{\bf 28  31  27 33  28 31 } & {\bf2.504975  }  &   {\bf15870 }  &  {\bf 1}$\lambda_6$    \\ 
       & 1.782835 & 196  217  189 231  197 216  & 2.505171   &   521556   &   7$\lambda_6$    \\ 
       & 1.846869 & 140  155  135 165  141 154  & 2.505244   &   266100   &   5$\lambda_6$    \\ 
       & 1.902901 & 112  124  108 132  113 123  & 2.505305   &   170304   &   4$\lambda_6$    \\ 
       & 1.942926 & 196  217  189 231  198 215  & 2.505347   &   521556   &   7$\lambda_6$    \\ 
       & 1.996294 & 84  93  81 99  85 92  & 2.505401   &   95796   &   3$\lambda_6$    \\ 
       & 2.071014 & 140  155  135 165  142 153  & 2.505474   &   266100   &   5$\lambda_6$    \\ 
       & 2.183101 & 56  62  54 66  57 61  & 2.505575   &   42576   &   2$\lambda_6$    \\ 
 \hline
\end{tabular}
}
\end{minipage}%
\caption{6 indices as defined in Ref.~[\onlinecite{PhysRevB.106.115410}] and [\onlinecite{Hermann:2012dy}] to get the commensurate simulation cells for the t3BN and BN/G/BN systems are listed above. We first find the 6 indices for a double-moir\'e \textit{commensurate} system, indicated by the bold letters, where two moir\'e interfaces have the same moir\'e length of $\lambda_{\ell}$: $\lambda_1 = 132.379~\AA$, $\lambda_2 =93.3206~\AA$, and $\lambda_3 = 71.6226~\AA$ for the t3BN with the twist angles of $\theta_{12}=\theta_{32}=1.08^{\circ},\ 1.54^{\circ}$, and $2.0^{\circ}$, and $\lambda_4 = 121.599~\AA$, $\lambda_5 =95.1874~\AA$, and $\lambda_6 = 128.048~\AA$ for the BN/G/BN (or BN/G/NB) with the $\theta_{12}=\theta_{32}=0.54^{\circ},\ 1.08^{\circ}$, and $1.62^{\circ}$.     
Fixing the twist angle $\theta_{12}$ between the first two layers using the same ratio of $(i,j,i^{\prime},j^{\prime})$, we find several \textit{incommnsurate} $\theta_{32}$, which satisfies the 2\% of the lattice tolerance strained from the $a_G$ or $a_{BN}$. The 4th column provides the strained lattice constant $|\bm{a}_1^{(3)}|$ for the top layer, and the last column contains the length of the commensurate simulation cell $\lambda$ as the multiples of the \textit{commensurate} $\lambda_{\ell}$ values.
}
\label{HermannIndices1}
\end{table*}
\begin{table*}[]
\begin{minipage}{.47\linewidth}
\resizebox{1.0\columnwidth}{!}{%
\begin{tabular}{|lccccc|}
\multicolumn{6}{c}{G/BN/G} \\
\hline
$\theta_{12}$ ($^{\circ}$) &$\theta_{32}$ ($^{\circ}$) & ($i$ $j$ $i'$ $j'$ $i''$ $j''$) & $|\bm{a}_1^{(3)}|$($\AA$) & \# atoms  & $\lambda$\\\hline
 0.601428 & 0.000000 & 56 54 54 54 55 55  & 2.460200   &   53798   &   2$\lambda_1$    \\ 
       & 0.085921 & 196 189 189 189 193 192  & 2.460197   &   659026   &   7$\lambda_1$    \\ 
       & 0.120290 & 140 135 135 135 138 137  & 2.460194   &   336238   &   5$\lambda_1$    \\ 
       & 0.200482 & 84 81 81 81 83 82  & 2.460185   &   121046   &   3$\lambda_1$    \\ 
       & 0.257762 & 196 189 189 189 194 191  & 2.460175   &   659030   &   7$\lambda_1$    \\ 
       & 0.300722 & 112 108 108 108 111 109  & 2.460166   &   215194   &   4$\lambda_1$    \\ 
       & 0.360865 & 140 135 135 135 139 136  & 2.460151   &   336242   &   5$\lambda_1$    \\ 
       & 0.400960 & 168 162 162 162 167 163  & 2.460140   &   484190   &   6$\lambda_1$    \\ 
       & 0.451078 & 224 216 216 216 223 217  & 2.460124   &   860786   &   8$\lambda_1$    \\ 
       & 0.515514 & 392 378 378 378 391 379  & 2.460100   &   2636174   &   14$\lambda_1$    \\ 
       & {\bf 0.601428} & {\bf 28 27 27 27 28 27}  & {\bf 2.460064 }  &   {\bf 13450 }  &   {\bf 1}$\lambda_1$    \\ 
       & 0.735065 & 252 243 243 243 253 242  & 2.459997   &   1089470   &   9$\lambda_1$    \\ 
       & 0.801881 & 168 162 162 162 169 161  & 2.459959   &   484214   &   6$\lambda_1$    \\ 
       & 0.841969 & 140 135 135 135 141 134  & 2.459934   &   336262   &   5$\lambda_1$    \\ 
       & 0.902100 & 112 108 108 108 113 107  & 2.459895   &   215210   &   4$\lambda_1$    \\ 
       & 0.945050 & 196 189 189 189 198 187  & 2.459865   &   659086   &   7$\lambda_1$    \\ 
       & 1.002314 & 84 81 81 81 85 80  & 2.459823   &   121058   &   3$\lambda_1$    \\ 
       & 1.082481 & 140 135 135 135 142 133  & 2.459761   &   336278   &   5$\lambda_1$    \\ 
       & 1.116837 & 196 189 189 189 199 186  & 2.459732   &   659110   &   7$\lambda_1$    \\ 
       & 1.202723 & 56 54 54 54 57 53        & 2.459658   &   53806   &   2$\lambda_1$    \\ 
       & 1.288603 & 196 189 189 189 200 185  & 2.459578   &   659138   &   7$\lambda_1$    \\ 
       & 1.322954 & 140 135 135 135 143 132  & 2.459544   &   336298   &   5$\lambda_1$    \\ 
       & 1.403102 & 84 81 81 81 86 79        & 2.459462   &   121070   &   3$\lambda_1$    \\ 
 \hline 
 1.276839 & 0.315623 & 63 279 54 279 57 283     & 2.459406   &   589358   &   9$\lambda_2$    \\ 
       & 0.411705    & 35 155 30 155 32 157     & 2.459539   &   181894   &   5$\lambda_2$    \\ 
       & 0.490326    & 77 341 66 341 71 345     & 2.459642   &   880342   &   11$\lambda_2$    \\ 
       & 0.555848    & 42 186 36 186 39 188     & 2.459725   &   261914   &   6$\lambda_2$    \\ 
       & 0.658820    & 49 217 42 217 46 219     & 2.459848   &   356482   &   7$\lambda_2$    \\ 
       & 0.736056    & 56 248 48 248 53 250     & 2.459935   &   465598   &   8$\lambda_2$    \\ 
       & 0.796132    & 63 279 54 279 60 281     & 2.460000   &   589262   &   9$\lambda_2$    \\ 
       & 0.844195    & 70 310 60 310 67 312     & 2.460050   &   727474   &   10$\lambda_2$    \\ 
       & 0.883521    & 77 341 66 341 74 343     & 2.460089   &   880234   &   11$\lambda_2$    \\ 
       & 0.916293    & 84 372 72 372 81 374     & 2.460121   &   1047542   &   12$\lambda_2$    \\ 
       & 0.944024    & 91 403 78 403 88 405     & 2.460148   &   1229398   &   13$\lambda_2$    \\ 
       & 0.988395    & 105 465 90 465 102 467   & 2.460189   &   1636754   &   15$\lambda_2$    \\ 
       & 1.006421    & 112 496 96 496 109 498   & 2.460205   &   1862254   &   16$\lambda_2$    \\ 
       & 1.212836    & 168 744 144 744 167 745  & 2.459417   &   4190978   &   24$\lambda_2$    \\ 
       & {\bf 1.276839} & {\bf 7 31 6 31 7 31}  & {\bf 2.460421 }  &   {\bf 7274 }  &   {\bf 1}$\lambda_2$    \\ 
       & 1.319537 & 252 1116 216 1116  253 1115  & 2.461089   &   9425378   &   36$\lambda_2$    \\ 
       & 1.637459 & 84 372 72 372 87 370         & 2.460623   &   1047398   &   12$\lambda_2$    \\ 
       & 1.709590 & 70 310 60 310 73 308         & 2.460651   &   727354   &   10$\lambda_2$    \\ 
       & 1.757677 & 63 279 54 279 66 277         & 2.460668   &   589154   &   9$\lambda_2$    \\ 
       & 1.817788 & 56 248 48 248 59 246         & 2.460687   &   465502   &   8$\lambda_2$    \\ 
       & 1.895074 & 49 217 42 217 52 215         & 2.460707   &   356398   &   7$\lambda_2$    \\ 
       & 1.998124 & 42 186 36 186 45 184         & 2.460727   &   261842   &   6$\lambda_2$    \\ 
       \hline 
 1.782679 & 1.276493 & 203 189 189 196 200 192  & 2.460978   &   683484   &   7$\lambda_3$    \\ 
          & 1.310243 & 145 135 135 140 143 137  & 2.460961   &   348718   &   5$\lambda_3$    \\ 
          & 1.388991 & 87 81 81 84 86 82        & 2.460916   &   125540   &   3$\lambda_3$    \\ 
          & 1.445237 & 203 189 189 196 201 191  & 2.460882   &   683502   &   7$\lambda_3$    \\ 
          & 1.487421 & 116 108 108 112 115 109  & 2.460855   &   223186   &   4$\lambda_3$    \\ 
          & 1.546477 & 145 135 135 140 144 136  & 2.460814   &   348732   &   5$\lambda_3$    \\ 
          & 1.613967 & 203 189 189 196 202 190  & 2.460765   &   683524   &   7$\lambda_3$    \\ 
          & 1.635057 & 232 216 216 224 231 217  & 2.460748   &   892770   &   8$\lambda_3$    \\ 
          & 1.698326 & 406 378 378 392 405 379  & 2.460698   &   2734146   &   14$\lambda_3$    \\ 
          & {\bf 1.782679} & {\bf 29 27 27 28 29 27 } & {\bf 2.460626 }  &   {\bf 13950}   &  {\bf  1}$\lambda_3$    \\ 
          & 1.867028 & 406 378 378 392 407 377  & 2.460548   &   2734258   &   14$\lambda_3$    \\ 
          & 1.951371 & 203 189 189 196 204 188  & 2.460466   &   683580   &   7$\lambda_3$    \\ 
          & 1.979484 & 174 162 162 168 175 161  & 2.460437   &   502226   &   6$\lambda_3$    \\ 
          & 2.018841 & 145 135 135 140 146 134  & 2.460396   &   348772   &   5$\lambda_3$    \\ 
          & 2.077875 & 116 108 108 112 117 107  & 2.460332   &   223218   &   4$\lambda_3$    \\ 
          & 2.176256 & 87 81 81 84 88 80        & 2.460219   &   125564   &   3$\lambda_3$    \\ 
          & 2.254955 & 145 135 135 140 147 133  & 2.460124   &   348798   &   5$\lambda_3$    \\ 
          & 2.372991 & 58 54 54 56 59 53        & 2.459973   &   55810   &   2$\lambda_3$    \\ \hline
\end{tabular}
}
\end{minipage}%
\begin{minipage}{.52\linewidth}
\centering
\resizebox{1.0\columnwidth}{!}{%
\begin{tabular}{|lccccc|}
\multicolumn{6}{c}{G/BN/BN Type I} \\
\hline
$\theta_{12}$ ($^{\circ}$) &$\theta_{32}$ ($^{\circ}$) & ($i$ $j$ $i'$ $j'$ $i''$ $j''$) & $|\bm{a}_1^{(3)}|$($\AA$) & \# atoms  & $\lambda$\\\hline
 1.202325 & -0.010739 & 58 52 56 54 57 55  & 2.461018   &   55138   &   2$\lambda_4$    \\ 
       & 0.107388 & 145  130  140 135  143 137  & 2.460961   &   344618   &   5$\lambda_4$    \\ 
       & 0.186135 & 87  78  84 81  86 82  & 2.460916   &   124064   &   3$\lambda_4$    \\ 
       & 0.284566 & 116  104  112 108  115 109  & 2.460855   &   220562   &   4$\lambda_4$    \\ 
       & 0.343622 & 145  130  140 135  144 136  & 2.460814   &   344632   &   5$\lambda_4$    \\ 
       & 0.411112 & 203  182  196 189  202 190  & 2.460765   &   675488   &   7$\lambda_4$    \\ 
       & 0.461728 & 290  260  280 270  289 271  & 2.460725   &   1378562   &   10$\lambda_4$    \\ 
       & 0.501094 & 435  390  420 405  434 406  & 2.460693   &   3101792   &   15$\lambda_4$    \\ 
       & {\bf0.579824 }& {\bf29  26  28 27  29 27}  & {\bf2.460626}   &   {\bf13786 }  &   {\bf1}$\lambda_4$    \\ 
       & 0.658550 & 435  390  420 405  436 404  & 2.460554   &   3101912   &   15$\lambda_4$    \\ 
       & 0.697911 & 290  260  280 270  291 269  & 2.460516   &   1378642   &   10$\lambda_4$    \\ 
       & 0.748516 & 203  182  196 189  204 188  & 2.460466   &   675544   &   7$\lambda_4$    \\ 
       & 0.815986 & 145  130  140 135  146 134  & 2.460396   &   344672   &   5$\lambda_4$    \\ 
       & 0.875019 & 116  104  112 108  117 107  & 2.460332   &   220594   &   4$\lambda_4$    \\ 
       & 0.973401 & 87  78  84 81  88 80  & 2.460219   &   124088   &   3$\lambda_4$    \\ 
       & 1.052100 & 145  130  140 135  147 133  & 2.460124   &   344698   &   5$\lambda_4$    \\ 
       & 1.170136 & 58  52  56 54  59 53  & 2.459973   &   55154   &   2$\lambda_4$    \\ 
       & 1.288158 & 145  130  140 135  148 132  & 2.459811   &   344728   &   5$\lambda_4$    \\ 
       & 1.366830 & 87  78  84 81  89 79  & 2.459697   &   124106   &   3$\lambda_4$    \\ 
       & 1.465159 & 116  104  112 108  119 105  & 2.459548   &   220642   &   4$\lambda_4$    \\ 
       & 1.524151 & 145  130  140 135  149 131  & 2.459455   &   344762   &   5$\lambda_4$    \\ \hline 
\end{tabular}
}
\newline
\vspace*{0.1 cm}
\newline
\resizebox{1.0\columnwidth}{!}{%
\begin{tabular}{|lccccc|}
\multicolumn{6}{c}{G/BN/BN Type II} \\
\hline
$\theta_{12}$ ($^{\circ}$) &$\theta_{32}$ ($^{\circ}$) & ($i$ $j$ $i'$ $j'$ $i''$ $j''$) & $|\bm{a}_1^{(3)}|$($\AA$) & \# atoms  & $\lambda$\\\hline
$-$1.202855 & -0.010739 & 54 56 56 54 57 55  & 2.461018   &   55122   &   2$\lambda_5$    \\ 
       & 0.107388 & 135  140  140 135  143 137  & 2.460961   &   344518   &   5$\lambda_5$    \\ 
       & 0.186135 & 81  84  84 81  86 82  & 2.460916   &   124028   &   3$\lambda_5$    \\ 
       & 0.284566 & 108  112  112 108  115 109  & 2.460855   &   220498   &   4$\lambda_5$    \\ 
       & 0.343622 & 135  140  140 135  144 136  & 2.460814   &   344532   &   5$\lambda_5$    \\ 
       & 0.382991 & 162  168  168 162  173 163  & 2.460786   &   496130   &   6$\lambda_5$    \\ 
       & 0.411112 & 189  196  196 189  202 190  & 2.460765   &   675292   &   7$\lambda_5$    \\ 
       & 0.461728 & 270  280  280 270  289 271  & 2.460725   &   1378162   &   10$\lambda_5$    \\ 
       & 0.501094 & 405  420  420 405  434 406  & 2.460693   &   3100892   &   15$\lambda_5$    \\ 
       & {\bf 0.579824 }& {\bf27  28  28 27  29 27 } & {\bf2.460626}   &  {\bf 13782 }  &   {\bf 1}$\lambda_5$    \\ 
       & 0.658550 & 405  420  420 405  436 404  & 2.460554   &   3101012   &   15$\lambda_5$    \\ 
       & 0.697911 & 270  280  280 270  291 269  & 2.460516   &   1378242   &   10$\lambda_5$    \\ 
       & 0.748516 & 189  196  196 189  204 188  & 2.460466   &   675348   &   7$\lambda_5$    \\ 
       & 0.815986 & 135  140  140 135  146 134  & 2.460396   &   344572   &   5$\lambda_5$    \\ 
       & 0.875019 & 108  112  112 108  117 107  & 2.460332   &   220530   &   4$\lambda_5$    \\ 
       & 0.973401 & 81  84  84 81  88 80  & 2.460219   &   124052   &   3$\lambda_5$    \\ 
       & 1.052100 & 135  140  140 135  147 133  & 2.460124   &   344598   &   5$\lambda_5$    \\ 
       & 1.170136 & 54  56  56 54  59 53  & 2.459973   &   55138   &   2$\lambda_5$    \\ 
       & 1.288158 & 135  140  140 135  148 132  & 2.459811   &   344628   &   5$\lambda_5$    \\ 
       & 1.366830 & 81  84  84 81  89 79  & 2.459697   &   124070   &   3$\lambda_5$    \\ 
       & 1.465159 & 108  112  112 108  119 105  & 2.459548   &   220578   &   4$\lambda_5$    \\ 
       & 1.524151 & 135  140  140 135  149 131  & 2.459455   &   344662   &   5$\lambda_5$ \\ \hline
\end{tabular}
}
\end{minipage}%
\caption{6 indices as defined in Ref.~[\onlinecite{PhysRevB.106.115410}] and [\onlinecite{Hermann:2012dy}] to get the commensurate simulation cells for the G/BN/G, and G/BN/BN (or G/BN/NB) systems are listed above, as similar to those of t3BN and BN/G/BN in TABLE~\ref{HermannIndices1}.
The moir\'e length of the double-moir\'e \textit{commensurate} cells are $\lambda_1=117.183~\AA$, $\lambda_2=86.1850~\AA$, and $\lambda_3=119.359~\AA$ for the G/BN/G of the $\theta_{32}=0.6^{\circ}, 1.28^{\circ},$ and $1.78^{\circ}$, and $\lambda_4=119.359~\AA$, and $\lambda_5=119.359~\AA$ for the Type I and II of G/BN/BN (or G/BN/NB), denoted by the bold letters.
}
\label{HermannIndices2}
\end{table*}


\section{Stacking dependent energy variations}
\label{Energies}

\begin{table*}[]
\begin{minipage}{1.0\linewidth}
\resizebox{1.0\linewidth}{!}{%
\begin{tabular}{|cl|cccc|cccc|cccc|}
\multicolumn{14}{c}{t3BN} \\
 \hline
 \multicolumn{2}{|c}{($^{*}$ for MSB)}&\multicolumn{4}{|c}{$\theta_{12}=\theta_{32}=1.08^{\circ}$} &\multicolumn{4}{|c}{$\theta_{12}=\theta_{32}=1.53^{\circ}$} &\multicolumn{4}{|c|}{$\theta_{12}=\theta_{32}=2.00^{\circ}$} \\ \hline
\multicolumn{2}{|c|}{(eV/atom)}  & E$_\text{tot}$ & E$_\text{el}$& E$_{\text{IF}_{12}}$  & E$_{\text{IF}_{23}}$ & E$_\text{tot}$& E$_\text{el}$  & E$_{\text{IF}_{12}}$  & E$_\text{IF23}$ & E$_\text{tot}$  & E$_\text{el}$ & E$_{\text{IF}_{12}}$ & E$_{\text{IF}_{23}}$ \\   \hline
BN/BN/BN &$\overline{\text{AAA}}^{*}$           & 
$-$6.715081 &$-$6.688514 &$-$0.013282 &$-$0.013286 & 
$-$6.714537 &$-$6.688618 &$-$0.012955 &$-$0.012965 & 
$-$6.714200 &$-$6.688807 &$-$0.012688 &$-$0.012704 \\
  &$\overline{\text{AAA}}$                & 
$-$6.715070 &$-$6.688528 &$-$0.013271 &$-$0.013271 & 
$-$6.714519 &$-$6.688648 &$-$0.012935 &$-$0.012935 & 
$-$6.714178 &$-$6.688857 &$-$0.012661 &$-$0.012661 \\ 
  &$\overline{\text{AAB}}$                & 
$-$6.714614 &$-$6.688527 &$-$0.013043 &$-$0.013044 & 
$-$6.714137 &$-$6.688853 &$-$0.012643 &$-$0.012641 & 
$-$6.713911 &$-$6.689080 &$-$0.012417 &$-$0.012414 \\
  &$\overline{\text{AAC}}$  & 
$-$6.714614 &$-$6.688527 &$-$0.013044 &$-$0.013043 & 
$-$6.714137 &$-$6.688853 &$-$0.012641 &$-$0.012643 & 
$-$6.713911 &$-$6.689080 &$-$0.012414 &$-$0.012417 \\ \hline
NB/BN/BN &  $\overline{\text{AAA}^{\prime}}$       & 
$-$6.714506 &$-$6.688474 &$-$0.013232 &$-$0.012800 & 
$-$6.713905 &$-$6.689061 &$-$0.012643 &$-$0.012201 & 
$-$6.713748 &$-$6.689216 &$-$0.012430 &$-$0.012102 \\
&$\overline{\text{AAB}^{\prime}}$       & 
$-$6.714451 &$-$6.688611 &$-$0.013151 &$-$0.012688 & 
$-$6.714015 &$-$6.688927 &$-$0.012745 &$-$0.012344 & 
$-$6.713811 &$-$6.689134 &$-$0.012497 &$-$0.012180 \\
  &$\overline{\text{AAC}^{\prime}}$       & 
$-$6.714750 &$-$6.688650 &$-$0.013293 &$-$0.012807 & 
$-$6.714279 &$-$6.688778 &$-$0.012956 &$-$0.012545 & 
$-$6.714001 &$-$6.688955 &$-$0.012690 &$-$0.012356 \\ \hline
BN/NB/BN &$\overline{\text{AA}^{\prime}\text{A}}^{*}$  & 
$-$6.714451 &$-$6.688757 &$-$0.012846 &$-$0.012848 & 
$-$6.714041 &$-$6.688925 &$-$0.012555 &$-$0.012561 & 
$-$6.713816 &$-$6.689093 &$-$0.012357 &$-$0.012365 \\ 
 & $\overline{\text{AA}^{\prime}\text{A}}$& 
$-$6.714447 &$-$6.688765 &$-$0.012841 &$-$0.012841 & 
$-$6.714034 &$-$6.688943 &$-$0.012545 &$-$0.012545 & 
$-$6.713809 &$-$6.689119 &$-$0.012345 &$-$0.012345 \\
  &$\overline{\text{AA}^{\prime}\text{B}}$& 
$-$6.714070 &$-$6.688896 &$-$0.012536 &$-$0.012638 & 
$-$6.713773 &$-$6.689141 &$-$0.012303 &$-$0.012328 & 
$-$6.713640 &$-$6.689279 &$-$0.012177 &$-$0.012184 \\
  &$\overline{\text{AA}^{\prime}\text{C}}$& 
$-$6.714070 &$-$6.688896 &$-$0.012638 &$-$0.012536 & 
$-$6.713773 &$-$6.689141 &$-$0.012328 &$-$0.012303 & 
$-$6.713640 &$-$6.689279 &$-$0.012184 &$-$0.012177 \\ \hline
  \multicolumn{13}{c}{} \\
\end{tabular}
}    
\end{minipage}
\begin{minipage}{1.0\linewidth}
\resizebox{1.0\linewidth}{!}{%
\begin{tabular}{|l|cccc|cccc|cccc|}
\multicolumn{13}{c}{G/BN/G} \\
 \hline
 &\multicolumn{4}{|c}{$\theta_{12}=\theta_{32}=0.60^{\circ}$} &\multicolumn{4}{|c}{$\theta_{12}=\theta_{32}=1.28^{\circ}$} &\multicolumn{4}{|c|}{$\theta_{12}=\theta_{32}=1.78^{\circ}$} \\ \hline
\multicolumn{1}{|c|}{(eV/atom)}  & E$_\text{tot}$ & E$_\text{el}$& E$_{\text{IF}_{12}}$  & E$_{\text{IF}_{23}}$ & E$_\text{tot}$& E$_\text{el}$  & E$_{\text{IF}_{12}}$  & E$_{\text{IF}_{23}}$ & E$_\text{tot}$  & E$_\text{el}$ & E$_{\text{IF}_{12}}$ & E$_{\text{IF}_{23}}$ \\   \hline
$\overline{\rm AAA}$  & $-$7.192579	&$-$7.151257	&$-$0.020661  &$-$0.020661   &$-$7.192415  & $-$7.151430 &	$-$0.020492 &	$-$0.020492&$-$7.192273	&$-$7.151619  & $-$0.020327 &$-$0.020327  \\ \hline
$\overline{\rm AAB}$  & $-$7.193096	&$-$7.150856	&$-$0.021142	&$-$0.021098 &$-$7.192624  & $-$7.151283 &	$-$0.020681 &	$-$0.020659 &$-$7.192357	&$-$7.151592  &	$-$0.020385	&$-$0.020378     \\ \hline
$\overline{\rm AAC}$  & $-$7.193096	&$-$7.150856	&$-$0.021098	&$-$0.021142 &$-$7.192624  & $-$7.151283		 &	$-$0.020659 &	$-$0.020681&$-$7.192357    &$-$7.151592  &	$-$0.020380& $-$0.020385    \\ \hline
\multicolumn{13}{c}{} \\
\end{tabular}
}    
\end{minipage}
\begin{minipage}{1.0\linewidth}
\resizebox{1.0\linewidth}{!}{%
\begin{tabular}{|cl|cccc|cccc|cccc|}
\multicolumn{14}{c}{BN/G/BN and BN/G/NB} \\
 \hline
 \multicolumn{2}{|c|}{}&\multicolumn{4}{|c}{$\theta_{12}=\theta_{32}=0.54^{\circ}$} &\multicolumn{4}{|c}{$\theta_{12}=\theta_{32}=1.08^{\circ}$} &\multicolumn{4}{|c|}{$\theta_{12}=\theta_{32}=1.62^{\circ}$} \\ \hline
\multicolumn{2}{|c|}{(eV/atom)}  & E$_\text{tot}$ & E$_\text{el}$& E$_{\text{IF}_{12}}$  & E$_{\text{IF}_{23}}$ & E$_\text{tot}$& E$_\text{el}$  & E$_{\text{IF}_{12}}$  & E$_{\text{IF}_{23}}$ & E$_\text{tot}$  & E$_\text{el}$ & E$_{\text{IF}_{12}}$& E$_{\text{IF}_{23}}$ \\   \hline
BN/G/BN &$\overline{\rm AAA}$  & $-$6.958184  & $-$6.915841 &	$-$0.021172 &	$-$0.021172 &		$-$6.958077 &	$-$6.915973 &	$-$0.021052 &	$-$0.021052 &		$-$6.957947 &	$-$6.916232 &	$-$0.020858 &	-0.020858  \\ 
&$\overline{\rm AAB}$  & $-$6.958819  & $-$6.915310 &	$-$0.021728 &	$-$0.021781 &		$-$6.958452 &	$-$6.915681 &	$-$0.021368 &	$-$0.021403 &		$-$6.958134 &	$-$6.916151 &	$-$0.020987 &	$-$0.020995     \\ 
&$\overline{\rm AAC}$  & $-$6.958819	 & $-$6.915310 &	$-$0.021781 &	$-$0.021728 &		$-$6.958452 &	$-$6.915679 &	$-$0.021404 &	$-$0.021369	&	$-$6.958134 &	$-$6.916151 &	$-$0.020996 &	$-$0.020987   \\ \hline
BN/G/NB &$\overline{\rm A^{\prime}AA}$  &  $-$6.958824 &	$-$6.915242 &	$-$0.021791 &	$-$0.021791 &		$-$6.958462 &	$-$6.915621 &	$-$0.021421 &	$-$0.021421 &		$-$6.958141 &	$-$6.916107 &	$-$0.021017 &	$-$0.021017\\ 
&$\overline{\rm A^{\prime}AB}$  &  $-$6.958536 &	$-$6.915112 &	$-$0.021699 &	$-$0.021725 &		$-$6.958149 &	$-$6.915775 &	$-$0.021185 &	$-$0.021189 &		$-$6.957986 &	$-$6.916110 &	$-$0.020938 &	$-$0.020938    \\ 
&$\overline{\rm A^{\prime}AC}$  &  $-$6.958814 &	$-$6.915371 &	$-$0.021722 &	$-$0.021722 &		$-$6.958443 &	$-$6.915734 &	$-$0.021355 &	$-$0.021354 &		$-$6.958127 &	$-$6.916191 &	$-$0.020968 &	$-$0.020968   \\ \hline
\end{tabular}
}    
\end{minipage}
\newline
\vspace*{0.22 cm}
\newline
\begin{minipage}{0.9\linewidth}
\resizebox{1.0\linewidth}{!}{%
\begin{tabular}{|cl|ccccc|ccccc|}
\multicolumn{12}{c}{G/BN/BN and G/BN/NB} \\
 \hline
\multicolumn{2}{|c}{ }&\multicolumn{5}{|c}{Type I} &\multicolumn{5}{|c|}{Type II} \\ 
\multicolumn{2}{|c}{ }&\multicolumn{5}{|c}{$\theta_{12}=+1.20^{\circ},\ \theta_{32}=+0.58^{\circ}$} &\multicolumn{5}{|c|}{$\theta_{12}=-1.20^{\circ},\ \theta_{32}=+0.58^{\circ}$} \\ \hline
\multicolumn{2}{|c|}{(eV/atom)}  & E$_\text{tot}$ & E$_\text{el}$& E$_{\text{IF}_{12}}$  & E$_{\text{IF}_{23}}$ & E$_{\text{IF}_\text{others}}$ & E$_\text{tot}$& E$_\text{el}$  & E$_{\text{IF}_{12}}$  & E$_{\text{IF}_{23}}$ & E$_{\text{IF}_\text{others}}$ \\   \hline
G/BN/BN &$\overline{\rm AAA}$ & 
$-$6.957145 &$-$6.929545 &$-$0.013035 &$-$0.014153 &$-$0.000413 &
$-$6.957084 &$-$6.929605 &$-$0.012997 &$-$0.014073 &$-$0.000408 \\
&$\overline{\rm AAB}$ & 
$-$6.957142 &$-$6.929519 &$-$0.013041 &$-$0.014170 &$-$0.000413 &
$-$6.957070 &$-$6.929585 &$-$0.013001 &$-$0.014076 &$-$0.000408 \\
&$\overline{\rm AAC}$ & 
$-$6.956907 &$-$6.929694 &$-$0.012854 &$-$0.013962 &$-$0.000397 & 
$-$6.957287 &$-$6.929617 &$-$0.013055 &$-$0.014203 &$-$0.000412 \\ \hline
G/BN/NB &$\overline{\rm AAA}$ & 
$-$6.956828 &$-$6.929714 &$-$0.012552 &$-$0.014170 &$-$0.000392 &
$-$6.956990 &$-$6.929764 &$-$0.012609 &$-$0.014225 &$-$0.000393 \\
&$\overline{\rm AAB}$ & 
$-$6.956886 &$-$6.929680 &$-$0.012597 &$-$0.014213 &$-$0.000396 &
$-$6.956740 &$-$6.929834 &$-$0.012443 &$-$0.014076 &$-$0.000387 \\
&$\overline{\rm AAC}$ & 
$-$6.956674 &$-$6.929837 &$-$0.012421 &$-$0.014035 &$-$0.000381 &
$-$6.956857 &$-$6.929736 &$-$0.012587 &$-$0.014143 &$-$0.000391 \\ \hline
\end{tabular}
}
\end{minipage}
\caption{
Top layer sliding and mirror-symmetry dependent elastic and interface energy configurations. We summarize the total, elastic, and interface potential energies of the \textit{commensurate} t3BN, G/BN/G, BN/G/BN, BN/G/NB, G/BN/BN, and G/BN/NB  
systems in eV/atom units. 
The sum of the interface energies (E$_{\text{IF}_{12}}$, E$_{\text{IF}_{23}}$, and E$_{\text{IF}_\text{others}}$) gives the potential energy (E$_{\text{pot}}$). 
The expressions for the different energy contributions are given in Eqs.~(\ref{eq:tot}),~(\ref{eq:el_pot}) and~(\ref{eq:el_pot2}). 
The asterisk(*) sign is used for the mirror symmetry broken (MSB) cases.
}
\label{tab:energy}
\end{table*}

\begin{equation}
\begin{aligned}
E_{\rm el} &= \frac{1}{2}\sum_{i} E_{\rm el}^{i} \\
E_{\rm pot} &= \frac{1}{2}\sum_{i} E_{\rm pot}^{i} 
            = E_{\rm IF_{12}}+E_{\rm IF_{23}}(+E_{\rm IF_{subs.}}) \\
E_{\rm IF_{\ell\ell^{\prime}}} &= \sum_{i\in\text{layer}\ell}\sum_{j\in\text{layer} \ell^{\prime}}
\phi_{\rm inter}^{ij} 
\quad (\text{for }\ell\neq\ell^{\prime})\\
E_{\rm IF_{subs.}} &= \sum_{i \in \text{subs.\ }} \sum_{j\notin\text{subs.\ }} \phi_{\rm inter}^{ij} \\
\end{aligned}
\label{eq:el_pot2}
\end{equation}

We present a summary of the total energy and other energy components for the full systems in TABLE~\ref{tab:energy}, as defined in Eqs.~(\ref{eq:tot}),~(\ref{eq:el_pot}) and~(\ref{eq:el_pot2}). 
In Eq.~(\ref{eq:el_pot2}), $E_{\rm IF_{\ell\ell^{\prime}}}$ represents the sum of interface potential energy terms between layers $\ell$ and $\ell^{\prime}$, while $E_{\rm IF_{subs.}}$ encompasses all interface energy terms originating from a substrate layer (subs.) when it is present, such as in G/BN/BN:NB and G/BN/NB:BN.  
Notably, the energy variations resulting from the sliding of the top layers are predominantly attributed to the interface potential energies, which are on the order of $0.5$ meV/atom. This magnitude is significantly larger (about 50 times) than that of the elastic energies, which are on the order of $0.01$ meV/atom. Additionally, we observe that in systems with mirror-symmetric properties, the breaking of mirror symmetry induced by bending corrugation further decreases the total energy by approximately $0.005$ meV/atom.

\section{Electrostatic effects}
\label{electrostatic}
In polar materials, including hBN, pairwise Coulomb interactions can also modify the final structure and energetics. 
In this section, based on benchmark calculations, we show that the local minima in the total energy at double-moir\'e \textit{commensuration} also persist in the presence of electrostatic effects.
\begin{figure*}[tbhp]
\begin{center}
\includegraphics[width=0.8\textwidth]{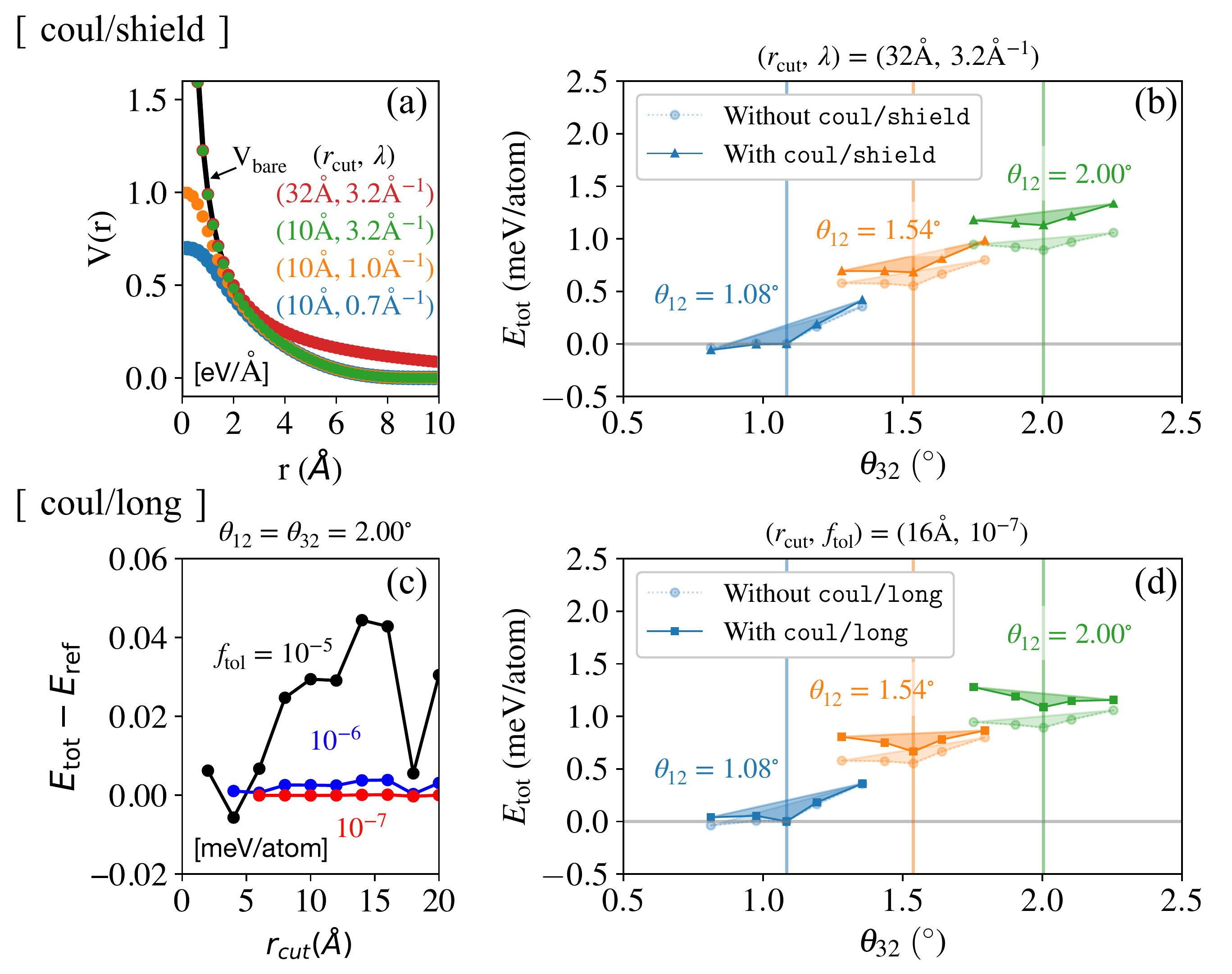}
\caption{(color online)
Electrostatic effects on the total energy per atom for t3BN ($\overline{\text{AAA}}$) using (top) \texttt{coul/shield} and (bottom) \texttt{coul/long} pair styles. 
(a) Dependence of the \texttt{coul/shield} potential on the cutoff distance $r_{\text{cut}}$ and shielding parameter $\lambda$: a tapering function smoothly brings the potential to zero at $r_{ij}=r_{\text{cut}}$, while $\lambda$ removes the $V_{\text{bare}}\propto 1/r_{ij}$ singularity at $r_{ij}=0$. 
(b) For $(r_{\text{cut}},\lambda)=(32\text{\AA},3.2\text{\AA}^{-1})$, chosen to mimic the bare Coulomb potential, local energy minima persist at the double-moir\'e \textit{commensurate} angles. 
(c) For \texttt{coul/long}, the total energy at $\theta_{12}=\theta_{32}=2.00^{\circ}$ is shown as a function of $r_{\text{cut}}$ for different force tolerances $f_{\text{tol}}=10^{-5},10^{-6},10^{-7}$, with the smallest fluctuations observed for $f_{\text{tol}}=10^{-7}$, where we therefore take $E_{\text{ref}}=-5.19356$ eV/atom as the converged reference energy.
(d) With \texttt{coul/long} at $(r_{\text{cut}},f_{\text{tol}})=(16\text{\AA},10^{-7})$, slightly deeper local energy dips appear at double-moir\'e \textit{commensuration}. 
Circles, triangles, and squares denote the cases without Coulomb interactions, with \texttt{coul/shield}, and with \texttt{coul/long}, respectively.
For the correction term $E_{\rm corr}$ in Eq.~(\ref{eq:tot}), $E_{L_3}^{\rm ref}$ is taken as $-6.690$~eV/atom for the first two cases and $-11.885$~eV/atom for the \texttt{coul/long} case. 
For ease of comparison, the energies per atom at $\theta_{12}=\theta_{32}=1.08^{\circ}$ are set to zero in each dataset.
}
\label{fig:energy_electrostatic}
\end{center}
\end{figure*}
In Fig. \ref{fig:energy_electrostatic}, we examine the total energies per atom as a function of $\theta_{32}$ when including the electrostatic pair styles \texttt{coul/shield} and \texttt{coul/long} in LAMMPS, focusing on the t3BN-$\overline{\text{AAA}}$ system.
Both pair styles implement Coulomb interaction potentials between atom pairs that are essentially proportional to $1/r_{ij}$, with modifications controlled by input parameters.
Panel (a) in Fig. \ref{fig:energy_electrostatic} shows how the \texttt{coul/shield} potential is controlled by the cutoff distance $r_{\text{cut}}$~\cite{coul_shield_Leven1, coul_shield_Leven2, coul_shield_Maaravi} and the shielding parameter $\lambda$~\cite{coul_shield_Maaravi}.
A tapering function smoothly decays the potential to zero at $r_{ij}=r_{\text{cut}}$, and the shielding parameter removes the $1/r$ singularity by replacing as $1/r_{ij}$ with $1/\sqrt[3]{r_{ij}^3+(1/\lambda)^3}$, which approaches $\lambda$ at $r_{ij}=0$. 
In panel (b), we present the total energy per atom in the presence of \texttt{coul/shield}, using a sufficiently large parameter set $(r_{\text{cut}},\ \lambda)=(32\text{\AA},3.2\text{\AA}^{-1})$ to mimic nearly bare Coulomb interactions.
The data with Coulomb effects (triangular markers) exhibit local minima at double-moir\'e \textit{commensurate} angles, with a steeper stabilization at smaller \textit{commensurate} angles, similar to the no-Coulomb case.
Panels (c) and (d) in Fig. \ref{fig:energy_electrostatic} illustrate the case of the \texttt{coul/long} pair style. 
Here, long-range Coulomb interactions are evaluated in reciprocal space, using the \texttt{pppm} k-space solver~\cite{coul_long_pppm_Hockney}. 
The numerical accuracy is controlled by the relative root-mean-square error in forces, $f_{\text{tol}}$, defined with respect to the force between two unit charges separated by $1\AA$.  
Panel (c) shows that the total energy per atom at $\theta_{12}=\theta_{32}=2.00^{\circ}$ as a function of the cutoff distance $r_{\text{cut}}$ for different values of $f_{\text{tol}}=10^{-5},10^{-6},10^{-7}$, with the reference energy set to $E_{\text{ref} }=-5.19356$ eV/atom.
We note that, in the \texttt{coul/long} case, $r_{\text{cut}}$ determines the real-space range beyond which the interaction is evaluated in reciprocal space, whereas in \texttt{coul/shield} case it defines the range beyond which the interaction is neglected.
Panel (d) shows similar energy-versus-$\theta_{32}$ trends in total energy per atom, but with slightly deeper local dips at the double-moir\'e \textit{commensuration}. Here we use $(r_{\text{cut}},f_{\text{tol}})=(16\AA,10^{-7})$. 
The slightly larger area between the dashed-line reference curve and the actual data points, together with the somewhat steeper slopes on the left and right of the \textit{commensurate} angles, suggests that long-range Coulomb interactions provide a modestly stronger energetic stabilization for these polar t3BN systems. 
%

\section{Comparison of Relaxation Effects in G/BN Bilayers}
\label{strain}
\begin{figure}[tbhp]
\begin{center}
\includegraphics[width=0.45\textwidth]{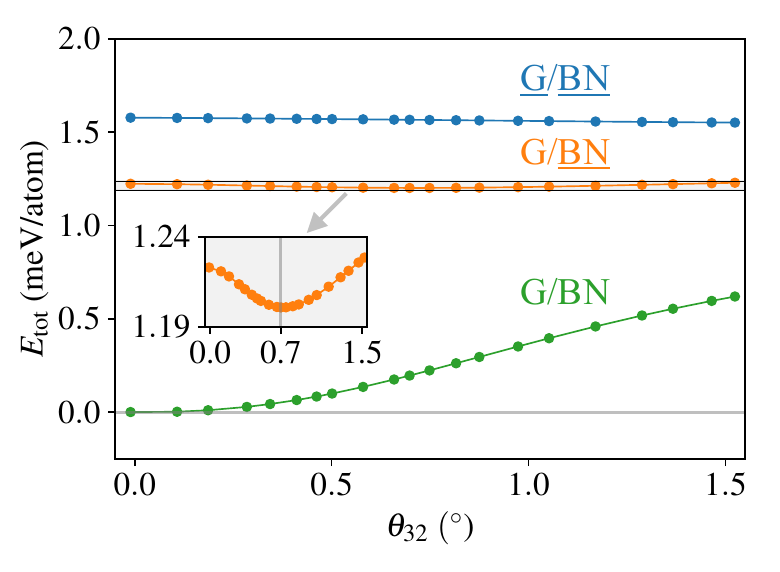}
\caption{(color online)
Total energy per atom of the rigid G/BN bilayer ($\underline{\mathrm{G}}/\underline{\mathrm{BN}}$), relaxed graphene on rigid hBN (G/$\underline{\mathrm{BN}}$), and fully relaxed G/BN bilayer as a function of twist angle $\theta_{32}$. 
The total energy in the $\underline{\mathrm{G}}/\underline{\mathrm{BN}}$ and G/$\underline{\mathrm{BN}}$ cases remains nearly constant, whereas that of the fully relaxed G/BN bilayer decreases toward its minimum as $\theta_{32}$ approaches zero. 
Inset shows the total-energy minimum near $\theta_{32}=0.7^{\circ}$ in the G/$\underline{\mathrm{BN}}$ case on a magnified energy scale and the energy difference relative to $0^{\circ}$ is about $0.02$~meV/atom, consistent with Ref.~\cite{GBN}. 
}
\label{fig:energyGBN_relaxation}
\end{center}
\end{figure}

To further investigate the origin of the global energy minimum in the G/BN/BN and G/BN/NB systems, we examine the total energy of the G/BN bilayer obtained by removing the bottom hBN layer from the corresponding double-moir\'e structures.
Fig.~\ref{fig:energyGBN_relaxation} shows the twist-angle dependence of the total energy for different relaxation conditions: fully rigid ($\underline{\mathrm{G}}/\underline{\mathrm{BN}}$), relaxed graphene on rigid hBN (G/$\underline{\mathrm{BN}}$), and fully relaxed (G/BN), where the underline denotes a rigid layer. 
The fully relaxed G/BN bilayer exhibits an energy minimum near zero twist angle, whereas the $\underline{\mathrm{G}}/\underline{\mathrm{BN}}$ and G/$\underline{\mathrm{BN}}$ cases show nearly constant total energies, consistent with the freestanding G/BN bilayer results reported in Ref.~\cite{GBN}.
Using the correction terms introduced in Sec.~\ref{methodSect}, we also recover the shallow total-energy minimum in the G/$\underline{\mathrm{BN}}$ case near $\theta_{32}=0.7^\circ$, as shown in the inset of Fig.~\ref{fig:energyGBN_relaxation}.
The total energy at the minimum is approximately 0.02 meV/atom lower than that at $\theta_{32}=0^\circ$.
Both the minimum angle and the corresponding energy difference are comparable to those reported in Ref.~\cite{GBN}.
However, the minimum angle obtained for the G/$\underline{\mathrm{BN}}$ case lies even farther from the double-moir\'e \textit{commensurate} angle than the $\sim0.6^\circ$ minimum reported in Ref.~\cite{GBN}.
These results support the interpretation that the global energy minimum in the G/BN/BN and G/BN/NB systems originates from double-moir\'e commensuration effects.
%


\newpage
\bibliography{multisystem.bib}

\end{document}